\documentclass{achemso}
\setkeys{acs}{usetitle=true}
\usepackage{amsmath}
\usepackage[version=3]{mhchem}
\usepackage{booktabs}
\usepackage{longtable}
\usepackage{multirow}
\usepackage{graphicx}
\usepackage{natbib}
\usepackage{xcolor}
\usepackage{float}
\usepackage[nolists,nomarkers]{endfloat}

\usepackage{xr-hyper}
\usepackage{hyperref}
\title[]{Predicting Novel Stable Materials for Experimental Synthesis}
\author{Yuqi An}
\affiliation[CityU]{Department of Materials Science and Engineering, City University of Hong Kong, Hong Kong SAR, 999077, China}
\author{Sihong Zhu}
\affiliation[CityU]{Department of Materials Science and Engineering, City University of Hong Kong, Hong Kong SAR, 999077, China}
\author{Joseph Montoya}
\affiliation[TRI]{Toyota Research Institute, Los Altos, CA 94022}
\author{Xingyu Guo}
\email{xingyguo@cityu.edu.hk}
\affiliation[CityU]{Department of Data Science, City University of Hong Kong, Hong Kong SAR, 999077, China}
\alsoaffiliation[CityU]{Hong Kong Institute of AI for Science, Hong Kong SAR, 999077, China}
\author{Zhenbin Wang}
\affiliation[CityU]{Department of Materials Science and Engineering, City University of Hong Kong, Hong Kong SAR, 999077, China}
\email{zwan22@cityu.edu.hk}
\date{}

\begin{document}

\maketitle

\begin{abstract}
Machine-learning-accelerated materials discovery has yielded large numbers of computationally stable compounds, yet many remain experimentally unrealized, underscoring a persistent gap between prediction and synthesis. Here, we introduce a hierarchical screening framework that combines PBE-based thermodynamic stability, efficient dynamical-stability screening enabled by universal machine-learning interatomic potentials, and SCAN-based thermodynamic refinement. Applying this protocol to the 894 stable materials previously reported in \textit{Sci. Data} 9, 302 (2022), we first curate 603 unique structures, of which only 298 remain thermodynamically stable on the complete PBE phase diagrams, demonstrating the critical role of competing phases in stability assessment. Dynamical screening then identifies 166 materials stable under both harmonic-phonon and finite-temperature molecular dynamics criteria, and SCAN phase diagrams further narrow this set to 109. Finally, by combining decomposition enthalpy with chemical-space completeness, we prioritize 25 candidates as high-confidence targets for experimental synthesis. This work provides a practical protocol for translating stability predictions into experimentally actionable synthesis targets, closing a key gap in machine-learning-driven materials discovery.

\end{abstract}

\section{Introduction}
The discovery of new materials with targeted properties is a major driver of technological innovation across diverse fields, including energy storage, catalysis, semiconductors, and quantum computing\cite{curtarolo2013high,butler2018machine}. Traditionally, new materials have been identified largely through experimental trial and error, a process that is time-consuming, resource-intensive, and often serendipitous. Computational materials prediction, particularly high-throughput density functional theory (DFT) calculations\cite{greeley2006computational,hautier2011phosphates,wang2018mining}, has transformed this paradigm by enabling the rapid screening of vast chemical spaces. More recently, advances in artificial intelligence have further accelerated materials discovery\cite{chen2022universal,merchant2023scaling,zeni2025generative,cavignac2025ai}. Deep learning models such as M3GNet\cite{chen2022universal} and GNoME\cite{merchant2023scaling} have identified millions of computationally stable compounds, achieving discovery rates orders of magnitude faster than conventional experimentation and offering the potential to substantially accelerate materials innovation.

Despite these significant advances, a critical challenge remains: most computationally predicted stable materials have yet to be experimentally synthesized. One important reason is that many of these materials may not be truly stable from an experimental standpoint\cite{cheetham2024artificial}. This disparity highlights a fundamental gap between computational prediction and experimental validation. Most theoretical predictions assess phase stability using DFT with the PBE exchange-correlation functional, typically based on criteria such as formation enthalpy\cite{bartel2022review}, and often supplement this with dynamical stability assessments from phonon spectra and/or short-timescale \textit{ab initio} molecular dynamics (AIMD) simulations \cite{malyi2019energy,xia2025search,ali2025exploring}. Although these approaches are effective for identifying thermodynamically favorable phases, they face two major limitations: limited predictive accuracy and high computational cost. On the accuracy front, the widely used PBE functional can yield unreliable phase-stability predictions, whereas the SCAN functional has been shown to substantially reduce errors in formation enthalpy calculations\cite{zhang2018efficient,wang2020predicting,an2025accelerating}, indicating that improved functionals can enhance the reliability of stability predictions. On the cost front, phonon calculations and AIMD simulations remain computationally demanding, and AIMD is typically restricted to picosecond timescales---too short to capture phase transitions that occur over longer periods. Recent developments in universal machine-learning interatomic potentials (uMLIPs) provide a promising solution by enabling efficient phonon calculations and long-timescale molecular dynamics simulations extending to nanoseconds, far beyond the practical reach of conventional AIMD\cite{chen2022universal,batatia2022mace,loew2025universal}. These capabilities offer near-DFT accuracy at a fraction of the computational cost, creating new opportunities for efficient and rigorous stability assessment.

In this work, we develop a hierarchical screening framework to improve the experimental relevance of computational stability predictions. We begin by revisiting 894 materials reported as stable in the Computational Autonomy for Materials Discovery (CAMD) dataset and reconstructing their PBE phase diagrams with an expanded set of competing phases. Candidates that remain thermodynamically stable are then evaluated using uMLIP-enabled harmonic phonon calculations and nanosecond-scale molecular dynamics simulations, which provide complementary tests of local 0~K dynamical stability and finite-temperature structural persistence. Finally, the dynamically robust candidates are refined using SCAN-calculated phase diagrams. This protocol reduces the original set of reported stable materials to 109 candidates that satisfy all three stability criteria: PBE thermodynamic stability, dynamical stability, and SCAN thermodynamic stability. Among these, 25 are prioritized as high-confidence synthesis targets by jointly considering decomposition enthalpy and chemical-space completeness. Beyond ranking candidates, this framework makes explicit why each predicted material warrants experimental investigation.

\section{Results}
\begin{figure}
    \centering
    \includegraphics[width=0.8\linewidth]{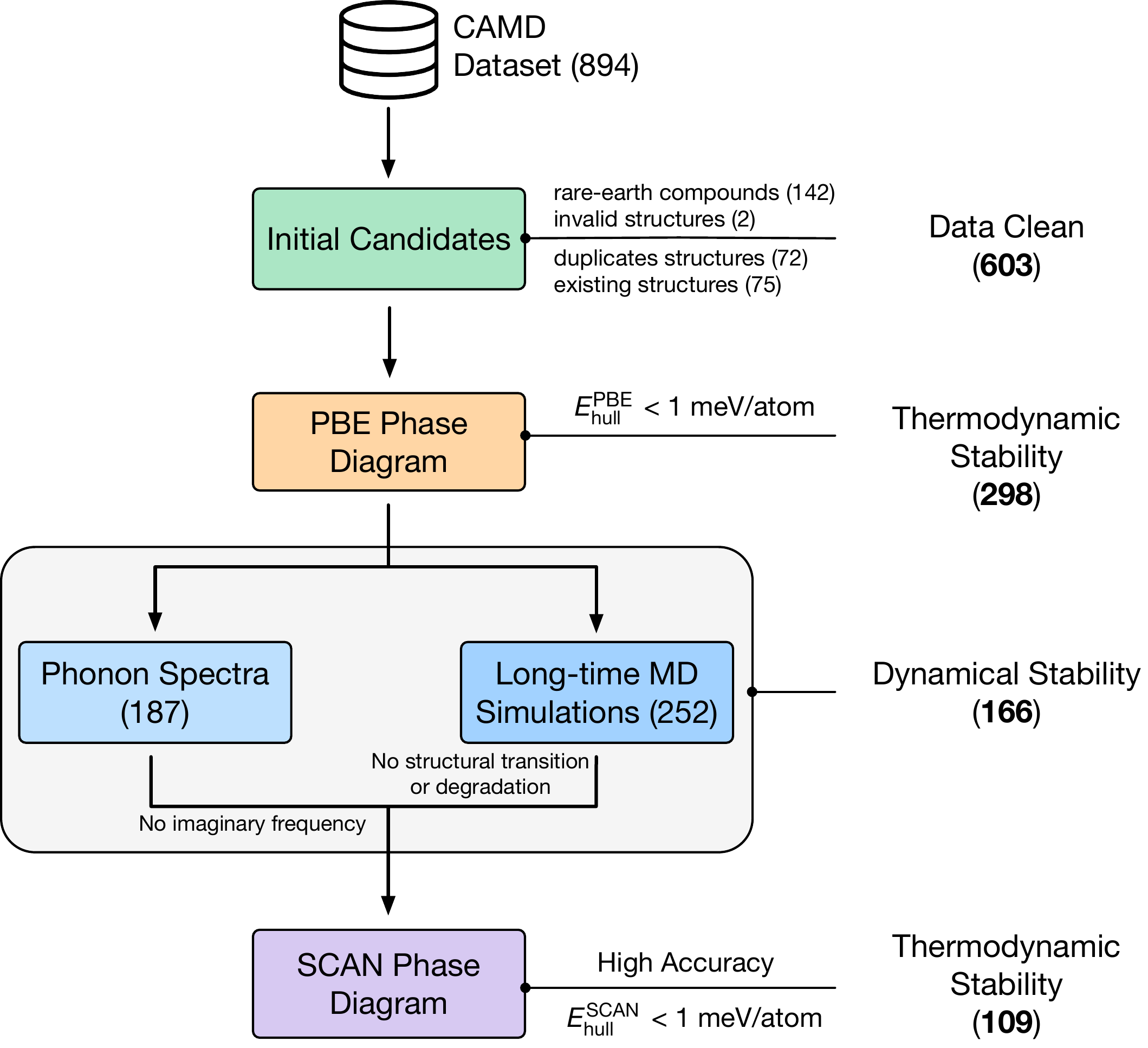}
    \caption{Schematic overview of the developed protocol for robust phase-stability prediction. Following initial data curation, the workflow comprises thermodynamic stability assessment at the PBE level, dynamical stability screening using harmonic phonon calculations and long-timescale molecular dynamics simulations, and final high-accuracy thermodynamic stability refinement using SCAN. The number of stable structures identified at each stage is given in parentheses.}
    \label{fig:flowchart}
\end{figure}

Figure~\ref{fig:flowchart} summarizes the three-stage screening protocol developed for robust phase-stability prediction of new materials. Following an initial data-curation step, the first stage screens the thermodynamic stability of the candidate structures using PBE-calculated phase diagrams, with the energy above the convex hull, $E_{\rm hull}$, used as the stability metric. Candidates predicted to be stable at the PBE level, defined here as $E^{\rm PBE}_{\rm hull} < 1$~meV/atom, are then subjected to dynamical-stability assessments. Two complementary tests are performed: harmonic phonon dispersion calculations at 0~K evaluate local dynamical stability from the absence of significant imaginary phonon modes, whereas long-timescale MD simulations at elevated temperatures are used to examine whether the crystal structures persist under finite-temperature atomic motion. The latter provides a trajectory-based test of structural persistence and naturally samples anharmonic effects beyond the harmonic approximation. In the final stage, candidates that pass both dynamical-stability tests are re-evaluated using SCAN-calculated phase diagrams to obtain a higher-fidelity assessment of their thermodynamic stability. This hierarchy identifies phases that are both thermodynamically and dynamically stable while reserving the more computationally demanding high-accuracy calculations for the most promising candidates.

\subsection{Dataset curation}

Starting from the CAMD dataset, we first excluded 142 compounds containing rare-earth elements. This exclusion was motivated by the substantial uncertainties often associated with DFT-calculated energetics for rare-earth systems, arising in part from the difficulty of treating localized $f$ electrons~\cite{eyring2002handbook} and selecting appropriate pseudopotentials~\cite{hegde2023quantifying,hautier_mp_psps} in a high-throughput setting. Two additional structures, \ce{Cs2Cd3O4} (camd-37758) and \ce{Rb3(NiO2)2} (camd-81553), were removed because their reported structures contain overlapping atomic coordinates. We then performed an all-against-all structure-matching analysis on the remaining 750 structures and identified 72 duplicates (Table~\ref{tab:duplicates}). For each duplicate group, the structure with the lowest $E^{\rm PBE}_{\rm hull}$ was retained for subsequent analysis; the energy differences among duplicates were negligible. Finally, we compared the remaining structures against the Materials Project database and found that 75 materials were already present, comprising 59 experimental structures and 16 Materials Project-calculated structures (Table~\ref{tab:matched_ss}). Overall, this data-cleaning procedure resulted in 603 unique candidate materials for subsequent phase-stability evaluation. These results underscore the importance of cross-checking predicted structures from high-throughput or machine-learning-based materials discovery against both the prediction dataset itself and established materials databases.

\subsection{PBE-based thermodynamic stability screening}

\begin{figure}[H]
    \centering
    \includegraphics[width=1.0\textwidth]{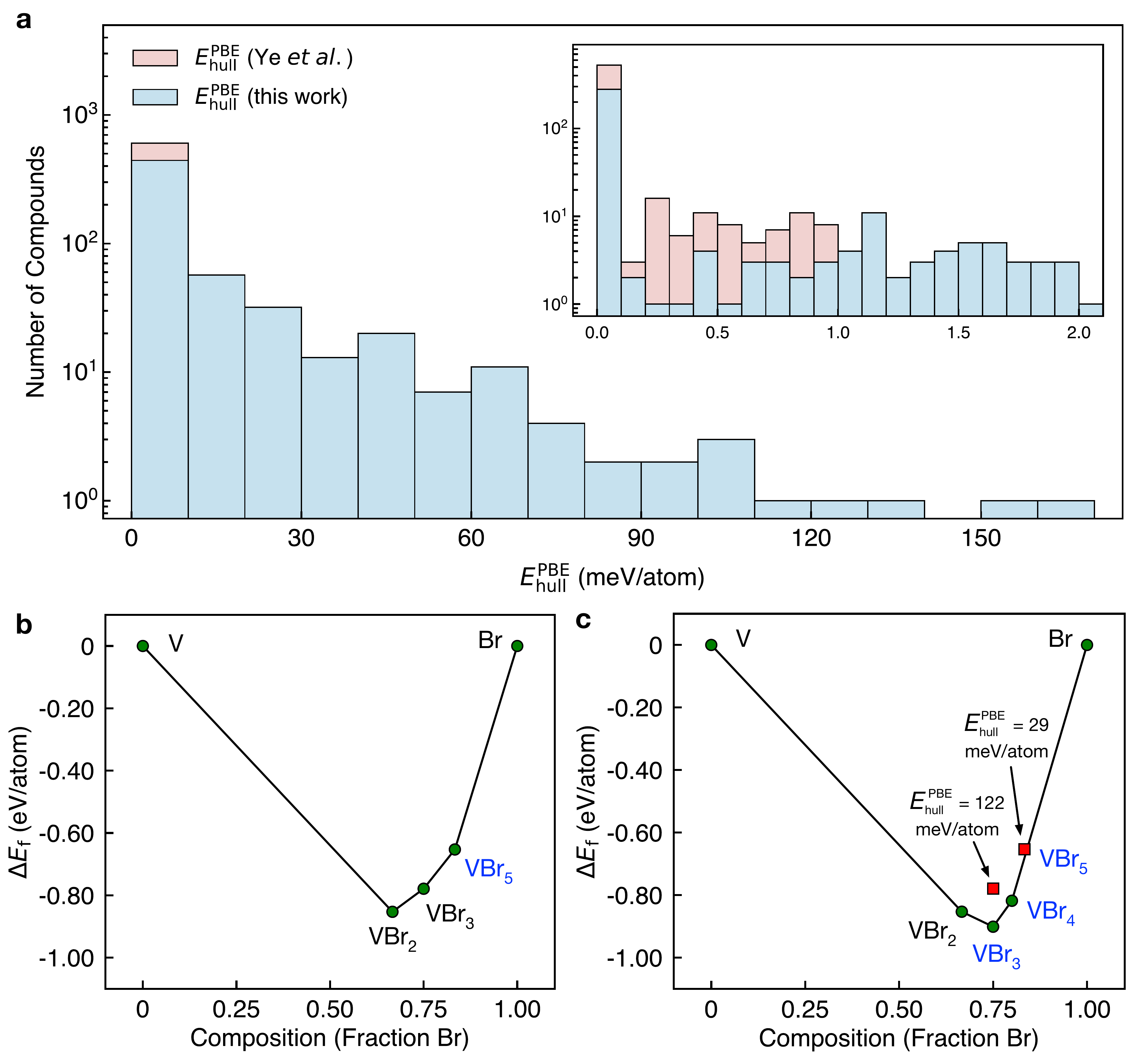}
    \caption{(a) Comparison of thermodynamic stabilities ($E^{\rm PBE}_{\rm hull}$) calculated using complete PBE phase diagrams for 603 unique materials with those reported in the CAMD dataset. The inset shows the distribution in the low-energy range of 0--2~meV/atom. (b) Calculated V--Br phase diagram constructed using only Materials Project entries together with the newly predicted \ce{VBr5} phase ($Cmcm$). (c) Recalculated V--Br phase diagram after inclusion of the additional competing CAMD phases \ce{VBr3} ($C2/c$) and \ce{VBr4} ($P\bar{4}3m$). Solid green circles denote thermodynamically stable phases on the convex hull, while the red square denotes a metastable phase. Compositions labeled in blue correspond to candidate materials from the CAMD dataset.}
    \label{fig:histogram}
\end{figure}

We next re-evaluated the thermodynamic stability of the 603 unique structures. By combining their calculated energies with those of competing phases drawn from the Materials Project, we constructed PBE-level phase diagrams and computed the corresponding energy above the convex hull, $E^{\rm PBE}_{\rm hull}$. Figure~\ref{fig:histogram}a compares our calculated $E^{\rm PBE}_{\rm hull}$ values with those reported by \citet{ye2022novel}. The recalculated $E^{\rm PBE}_{\rm hull}$ values range from 0 to 164~meV/atom, and only 298 materials (49.4\%) exhibit $E^{\rm PBE}_{\rm hull}$ below 1~meV/atom and are thus predicted to be thermodynamically stable (Table~\ref{tab:pbe_stables}). This finding contrasts with the report of \citet{ye2022novel}, in which all of these materials were classified as stable. The discrepancy arises from the set of competing phases included in the phase-diagram construction: the completeness of the sampled chemical space strongly affects the calculated $E_{\rm hull}$ and, consequently, the predicted stability of a candidate material.

The V--Br chemical system illustrates this effect. The Materials Project identifies \ce{VBr2} ($P\bar{3}m1$, mp-971787) and \ce{VBr3} ($P6_3/mmc$, mp-865473) as stable phases on the binary phase diagram, as shown in Figure~\ref{fig:histogram}b. Using only these Materials Project reference phases, the newly reported \ce{VBr5} structure ($Cmcm$, camd-16802) is predicted to be stable, with $E^{\rm PBE}_{\rm hull}=0$. However, when all V--Br compounds from the CAMD dataset are included in the phase-diagram construction, two additional stable phases, \ce{VBr3} ($C2/c$, camd-33441) and \ce{VBr4} ($P\bar{4}3m$, camd-32457), emerge on the convex hull (Figure~\ref{fig:histogram}c). The inclusion of these phases lowers the convex hull, rendering \ce{VBr5} metastable with a recalculated $E^{\rm PBE}_{\rm hull}$ of 29~meV/atom, and destabilizes the Materials Project \ce{VBr3} polymorph, which shifts to $E^{\rm PBE}_{\rm hull}=122$~meV/atom. These results demonstrate that a calculated $E_{\rm hull}$ is inherently dependent on the completeness of the competing phase space. An incomplete reference set can overestimate thermodynamic stability or, equivalently, underestimate the true $E_{\rm hull}$.

\subsection{Dynamical stability}

\begin{figure}[H]
\centering
\includegraphics[width=1.0\linewidth]{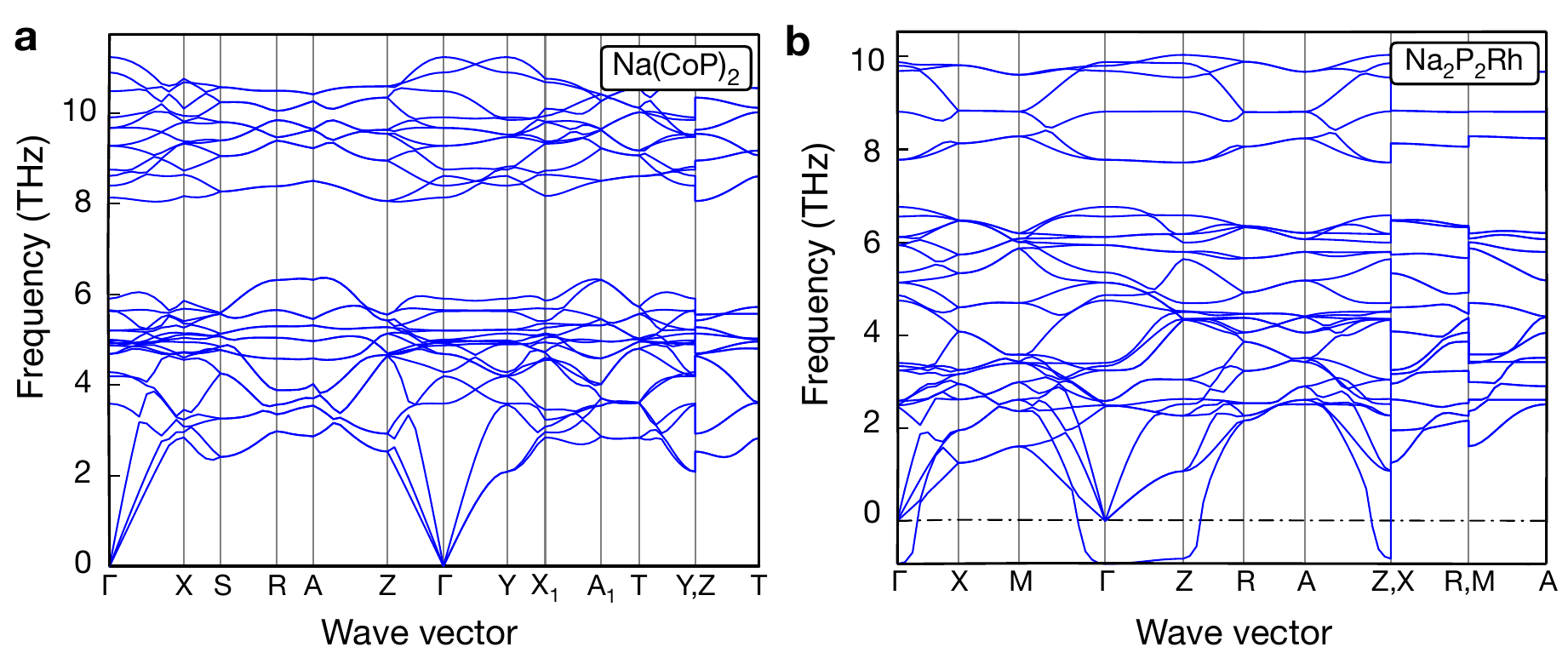}
\caption{Calculated phonon spectra for representative cases of dynamical stability and instability: (a) dynamically stable \ce{Na(CoP)2} ($I4/mmm$) and (b) dynamically unstable \ce{Na2P2Rh} ($P4/mbm$), in which imaginary phonon frequencies are observed.}
\label{fig:phonon-sample}
\end{figure}

To further evaluate the 298 thermodynamically stable candidates, we assessed their dynamical stability using both harmonic phonon calculations and long-timescale MD simulations. As noted above, phonon dispersions establish local harmonic dynamical stability at 0~K, whereas MD simulations probe finite-temperature structural robustness and capture anharmonic atomic motion. Both calculations are computationally demanding at the \textit{ab initio} level; consequently, they are rarely applied together in large-scale high-throughput studies and are typically restricted to a small number of candidate materials.

\subsubsection{Harmonic phonon stability}

We calculated harmonic phonon spectra for all 298 candidate materials using MACE. Structures free of imaginary modes were classified as dynamically stable at 0~K, whereas those exhibiting imaginary modes are classified as dynamically unstable. Representative spectra for each case are shown in Figure~\ref{fig:phonon-sample}. Applying this criterion, we found that 187 of the 298 materials are dynamically stable.

\begin{figure}[H]
\centering
\includegraphics[width=0.9\linewidth]{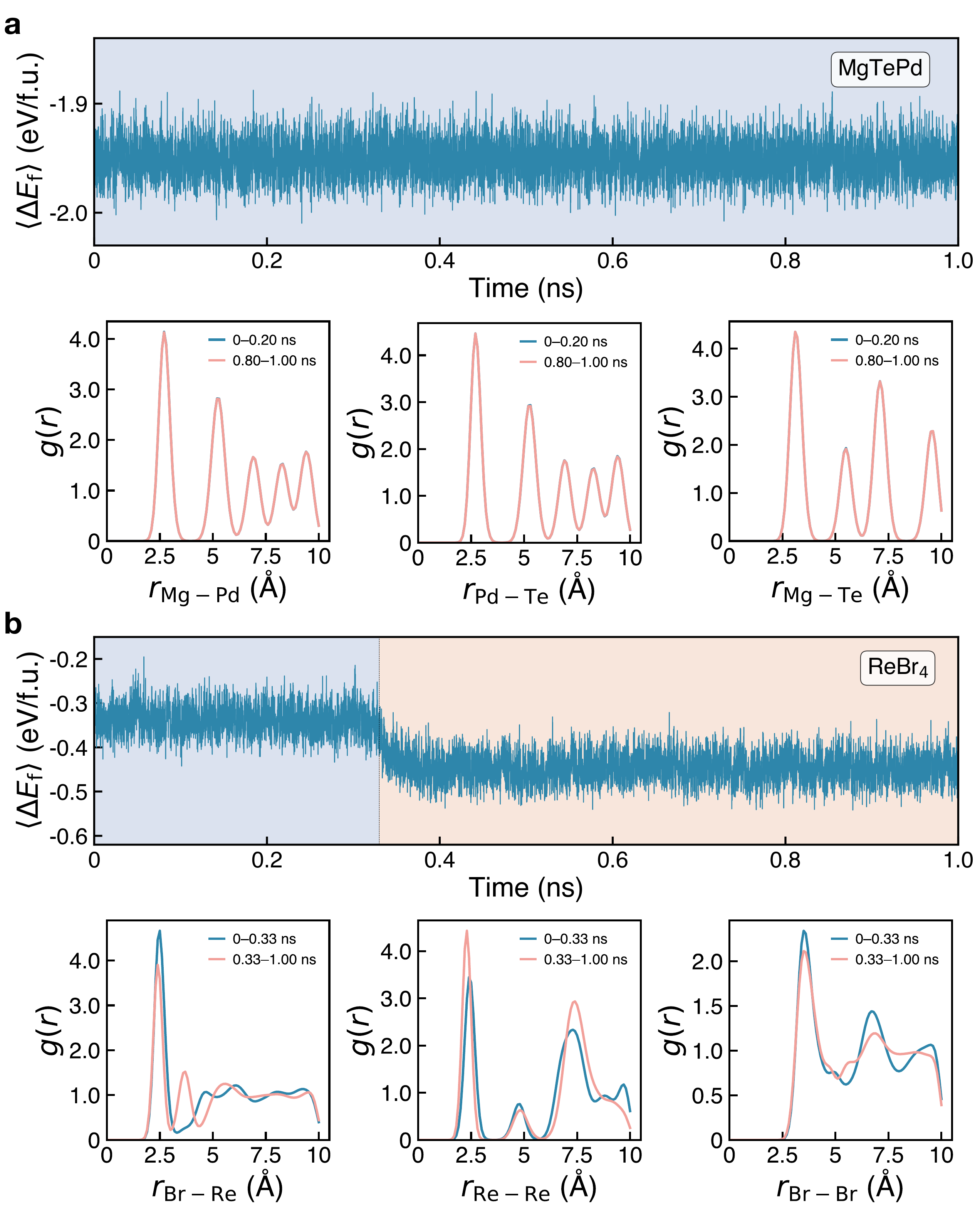}
\caption{Time evolution of the averaged formation potential energy ($\left\langle \Delta E_{\mathrm{f}} \right\rangle$) over 1~ns molecular dynamics simulations for (a) \ce{MgTePd} ($F\bar{4}3m$) and (b) \ce{ReBr4} ($P2/c$), with the corresponding radial distribution functions (RDFs) shown before and after the identified change point in $\left\langle \Delta E_{\mathrm{f}} \right\rangle$. For \ce{MgTePd}, no change point is detected; the RDFs are therefore averaged over the first and last 0.2~ns of the trajectory for comparison.}
\label{fig:md_sample_figure}
\end{figure}

\begin{figure}[H]
\centering
\includegraphics[width=0.9\linewidth]{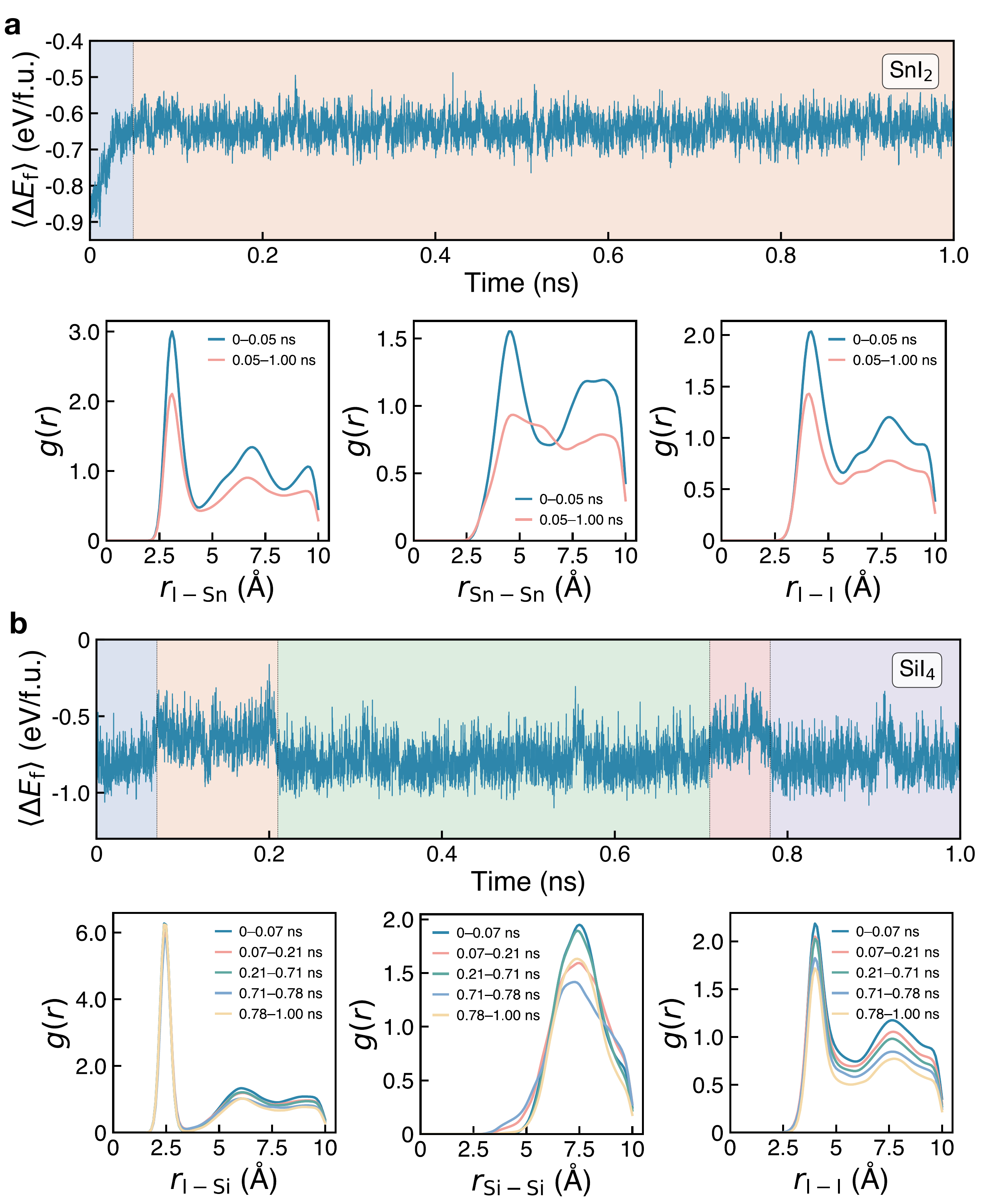}
\caption{Time evolution of the averaged formation potential energy ($\left\langle \Delta E_{\mathrm{f}} \right\rangle$) over 1~ns MD simulations of (a) \ce{SnI2} ($P\bar{1}$) and (b) \ce{SiI4} ($I\bar{4}2m$), with the corresponding radial distribution functions (RDFs) computed before and after the detected energy change point.}
\label{fig:md_sample_figure-2}
\end{figure}

\subsubsection{Finite-temperature MD stability}

To complement the harmonic phonon analysis, we examined the finite-temperature structural stability of the 298 materials using nanosecond-scale MD simulations. Compared with short AIMD simulations on the order of tens of picoseconds\cite{malyi2019energy,xia2025search,ali2025exploring}, these long-timescale simulations provide a more stringent assessment of structural persistence under sustained thermal motion while also capturing anharmonic effects. Structural stability along each trajectory was assessed using two complementary descriptors: the formation energy ($\Delta E_{\mathrm{f}}$), which monitors energetic changes associated with possible structural degradation or phase transformation, and the radial distribution function (RDF), which captures changes in local structural order during the simulation. To avoid subjective visual inspection, change points in the formation-energy trajectories were detected automatically using the pruned exact linear time (PELT) algorithm.\cite{TRUONG2020107299} 

Long-timescale MD simulations identified 252 of the 298 candidate materials as dynamically stable over the course of the 1~ns trajectories. To interpret the finite-temperature behavior across all candidates, we analyzed the time evolution of the formation energy together with RDF-based structural changes. Based on these energetic and structural descriptors, the MD trajectories can be grouped into four representative stability regimes, as illustrated in Figure~\ref{fig:md_sample_figure} and Figure~\ref{fig:md_sample_figure-2}.

In the first regime, the materials remain kinetically stable throughout the simulation. As shown for \ce{MgTePd} in Figure~\ref{fig:md_sample_figure}a, the formation energy fluctuates around a nearly constant mean value, with no discernible drift or discontinuity over the full 1~ns trajectory. The RDF remains essentially unchanged, with a small difference between the early and late trajectory windows ($\Delta$RDF $<0.01$). These results indicate that the crystalline structure is preserved under sustained thermal motion throughout the MD simulation.

The second regime corresponds to metastable structures that relax into lower-energy configurations during MD. For example, \ce{ReBr4} remains near its initial formation-energy plateau for the first $\sim$0.33~ns, followed by a pronounced energy decrease between 0.34 and 0.40~ns (Figure~\ref{fig:md_sample_figure}b). This transition is accompanied by substantial RDF changes, with $\Delta \mathrm{RDF}_{\mathrm{Br-Re}} = 0.22$ and $\Delta \mathrm{RDF}_{\mathrm{Re-Re}} = 0.28$, indicating a significant structural rearrangement. This behavior suggests that the 0~K DFT-optimized structure is a metastable phase rather than the lowest-energy accessible state. Thermal fluctuations during MD allow the system to overcome kinetic barriers and relax into a more favorable phase that may be missed by static structure optimization alone.

The third regime involves thermally induced structural transformations. As exemplified by \ce{SnI2} in Figure~\ref{fig:md_sample_figure-2}a, the formation energy increases rapidly at the beginning of the simulation and then reaches a higher-energy plateau. At the same time, the RDF changes substantially for all atomic pair correlations, with $\Delta$RDF values of approximately 0.2. This behavior indicates that a structure that is stable at 0~K may lose its original structural order at elevated temperature and transform into a distinct finite-temperature configuration.

The fourth regime is characterized by structural fragility. As shown for \ce{SiI4} in Figure~\ref{fig:md_sample_figure-2}b, the formation-energy trajectory exhibits frequent large-amplitude fluctuations and abrupt transitions between multiple states, without reaching a well-defined plateau. Corresponding RDF changes at these transitions range from 0.01 to more than 0.1 for different atomic pair correlations, reflecting pronounced structural fluxionality and possible loss of crystalline order. These materials are therefore unlikely to maintain a stable crystalline framework under finite-temperature conditions.

Combining the harmonic phonon analysis with the finite-temperature MD simulations, we identify 166 materials that are dynamically stable under both criteria. These materials exhibit no imaginary phonon modes at 0~K and retain their crystalline structures throughout 1~ns of thermal motion, thereby constituting the most robust subset of the 298 candidates for experimental synthesis and further property evaluation. The full list is provided in Table~\ref{tab:dynamically_stable}.

The remaining materials fall into three diagnostic categories that highlight the complementarity of the two stability criteria. Twenty-one materials are phonon-stable but MD-unstable, indicating finite-temperature instabilities or phase transitions driven by anharmonic effects beyond the harmonic approximation. Conversely, 86 materials are phonon-unstable but MD-stable, suggesting that the soft modes predicted at 0~K are either stabilized by finite-temperature anharmonicity or correspond to shallow local instabilities that do not trigger structural collapse on the MD timescale. Finally, 25 materials are unstable in both analyses. Notably, the two tests disagree for more than one-third of the candidates (107 of 298), demonstrating that harmonic phonon calculations and long-timescale MD provide complementary, non-redundant information. Thus, combining both criteria provides a more rigorous stability filter for identifying dynamically robust materials.

\subsection{SCAN-refined thermodynamic stability}

\begin{figure}[H]
    \centering
    \includegraphics[width=1.0\linewidth]{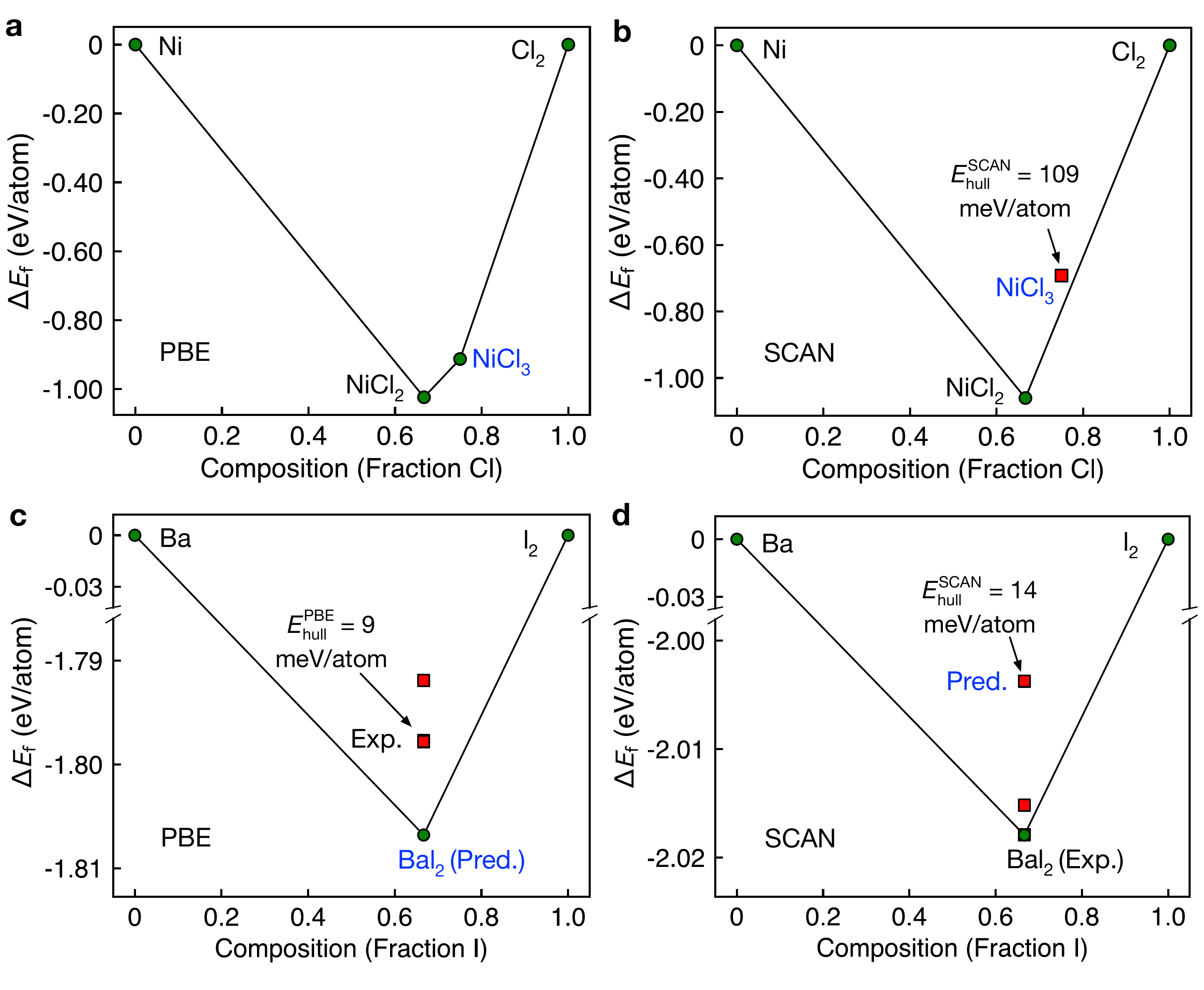}
    \caption{Comparison of calculated phase diagrams for the representative Ni--Cl and Ba--I chemical systems computed using the PBE and SCAN functionals, illustrating the effect of the exchange-correlation functional on the predicted $E_{\rm hull}$. Panels show (a) Ni--Cl with PBE, (b) Ni--Cl with SCAN, (c) Ba--I with PBE, and (d) Ba--I with SCAN. Solid green circles denote thermodynamically stable phases on the convex hull, while the red square denotes a metastable phase. Compositions labeled in blue correspond to candidate materials from the CAMD dataset. Labels ``Exp.'' and ``Pred.'' refer to experimentally reported and CAMD-predicted phases, respectively.}
    \label{fig:scan_pbe_pd}
\end{figure}

Previous studies\cite{zhang2018efficient,wang2020predicting} have shown that the SCAN functional approximately halves the error in calculated formation energies relative to PBE. We therefore reassessed the thermodynamic stability of the 166 dynamically stable candidates using the energy above the convex hull computed from SCAN-derived phase diagrams, $E^{\rm SCAN}_{\rm hull}$. Of these candidates, 109 materials (65.7\%) have $E^{\rm SCAN}_{\rm hull}<1$~meV/atom, supporting their assignment as stable phases.

To illustrate the impact of this refinement, we highlight two representative binary systems in Figure~\ref{fig:scan_pbe_pd}. Figures~\ref{fig:scan_pbe_pd}a and \ref{fig:scan_pbe_pd}b compare the Ni--Cl phase diagrams obtained with PBE and SCAN, respectively. The \ce{NiCl3} phase with $R\bar{3}$ symmetry is predicted to be thermodynamically stable by PBE but is placed well above the convex hull by SCAN, with $E_{\mathrm{hull}}^{\mathrm{SCAN}}=109$~meV/atom. The SCAN prediction is consistent with the experimental Ni--Cl phase diagram, in which only \ce{NiCl2} has been observed as a stable nickel chloride phase\cite{okamoto2000phase}. This result is also chemically intuitive, as the strongly oxidizing \ce{Ni^{3+}} ion can oxidize \ce{Cl^-} anions, favoring decomposition of \ce{NiCl3} into \ce{NiCl2} and chlorine gas.

The second example concerns the polymorphism of \ce{BaI2}. In the PBE phase diagram, shown in Figure~\ref{fig:scan_pbe_pd}c, a hypothetical $Fm\bar{3}m$ phase is predicted to be the ground state, whereas the experimentally known $Pnma$ phase lies 9~meV/atom above the hull. SCAN reverses this ordering (Figure~\ref{fig:scan_pbe_pd}d), placing the experimental $Pnma$ phase as the ground state and the hypothetical $Fm\bar{3}m$ phase 14~meV/atom above the hull. Given that \ce{BaI2} is a simple and well-characterized binary compound\cite{okamoto2000phase}, the experimentally observed phase is very likely to be the true ground state, and only SCAN recovers this ordering. These examples demonstrate that higher-fidelity functionals such as SCAN can be critical for reliable phase-stability prediction.

\section{Discussion}

\begin{figure}[H]
    \centering
    \includegraphics[width=0.9\linewidth]{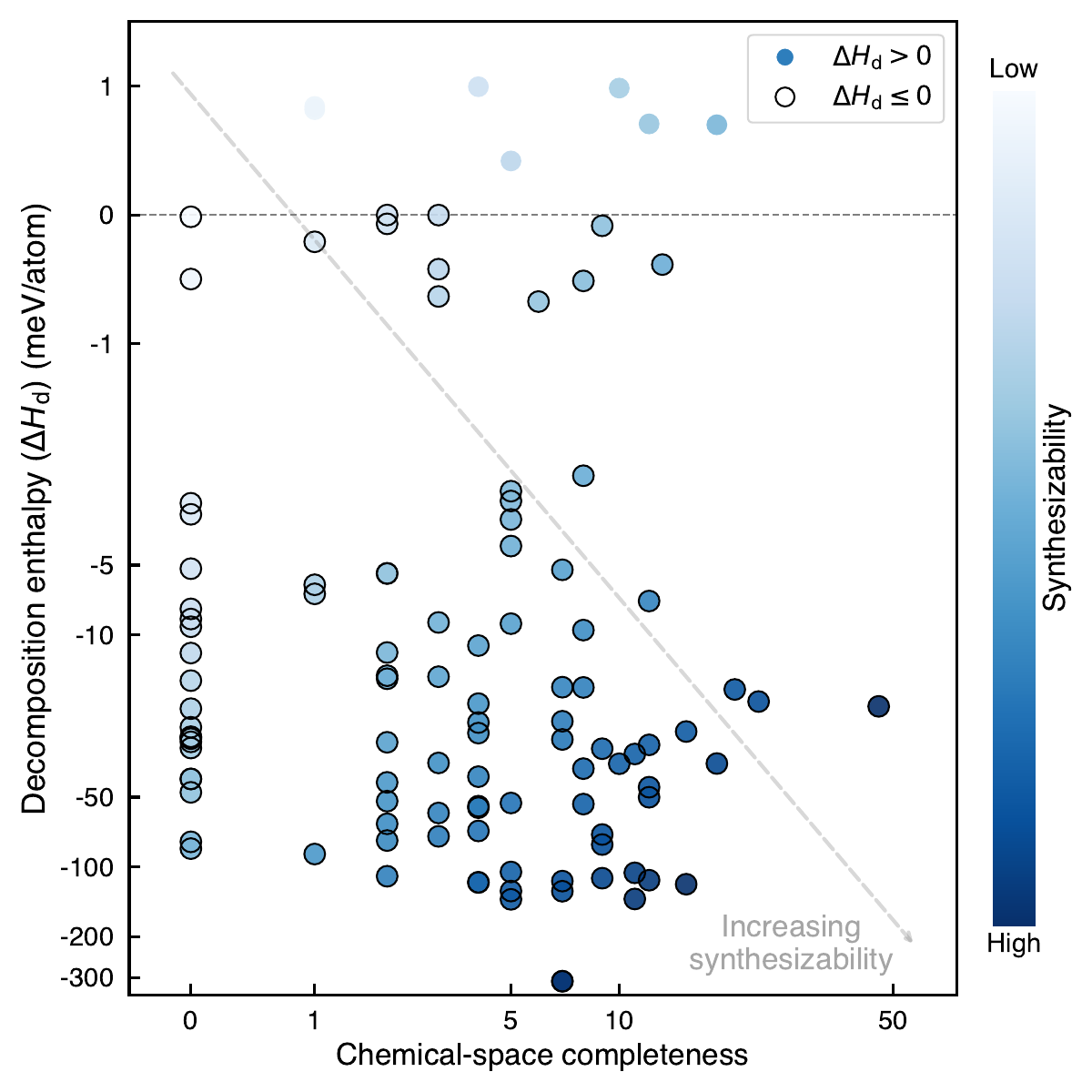}
    \caption{Distribution of decomposition enthalpy, $\Delta H_{\mathrm{d}}$, as a function of chemical-space completeness for the 109 candidate materials. Chemical-space completeness is quantified by the number of known phases, $C$, in the corresponding chemical system, with the $x$-axis plotted on a logarithmic scale. Marker color indicates the composite synthesizability score, which combines log-scaled chemical-space completeness and log-compressed thermodynamic stability; darker shades correspond to a higher predicted likelihood of synthesizability. Open circles denote structures with $\Delta H_{\mathrm{d}} \leq 0$ meV/atom. The dashed arrow indicates the direction of increasing synthesizability confidence, toward higher chemical-space completeness and lower decomposition enthalpy.}
    \label{fig:completeness_distribution}
\end{figure}

As discussed above, the reliability of hull-based stability metrics depends on the completeness of the competing phases sampled in a given chemical system. A candidate may appear stable or nearly stable simply because the chemical space is incompletely explored, and newly discovered phases could overturn its predicted stability. To account for this uncertainty, we evaluated the decomposition enthalpy ($\Delta H_{\rm d}$)\cite{bartel2019role} together with chemical-space completeness, defined here as the total number of known competing phases in the same chemical system. This combined descriptor provides a practical measure of confidence in predicted synthesizability. Specifically, we defined a composite synthesizability score, $S$, by equally weighting the normalized chemical-space-completeness and thermodynamic-stability contributions: $S=\operatorname{MinMax}(\tilde{C}+\tilde{E})$, where $\tilde{C}=\operatorname{MinMax}[\log_{10}(C+1)]$ and $\tilde{E}=1-\operatorname{MinMax}[\operatorname{sgn}(\Delta H_{\rm d})\log_{10}(1+|\Delta H_{\rm d}|)]$. Here, $C$ is the number of known competing phases in the corresponding chemical system. $S$ ranges from 0 to 1, with larger values indicating higher confidence in synthesizability.

Figures~\ref{fig:completeness_distribution} shows the distribution of $\Delta H_{\rm d}$ versus chemical-space completeness for the 109 candidate materials. To reduce computational cost, $\Delta H_{\rm d}$ was calculated using PBE, as previous work has shown that PBE and SCAN yield comparable values for this metric~\cite{bartel2019role}. Candidates in the lower-right region combine low decomposition enthalpy with well-explored chemical spaces, and their favorable stability is therefore less likely to be overturned by the discovery of new competing phases. On this basis, we identify 25 candidates ($S> 0.80$), corresponding to 23\% of the screened materials, as high-confidence targets for experimental synthesis. The composite synthesizability scores for all 109 materials are provided in Table~\ref{tab:S_scores}.

This study uses the CAMD dataset to develop a multi-stage screening protocol that combines thermodynamic and dynamical stability criteria, yielding 109 promising candidates for experimental validation. The use of uMLIPs substantially reduces the cost of phonon and MD calculations and enables stability assessments at length scales and timescales that are impractical with DFT alone, although the accuracy of uMLIP-based phonon predictions remains model- and chemistry-dependent.\cite{loew2025universal} One important limitation lies in the timescales accessible to MD simulations when evaluating finite-temperature structural stability. Phase transformations and structural degradation are often slow, thermally activated processes that short MD trajectories cannot fully capture. In the literature, such stability is commonly assessed using AIMD simulations lasting only tens of picoseconds, a time window over which many of these processes are unlikely to occur. The nanosecond-scale simulations performed here extend this window by roughly two orders of magnitude and therefore provide a more stringent test. Even so, the absence of structural degradation over nanoseconds is not sufficient to establish long-term stability, because some transformations may occur over timescales ranging from microseconds to macroscopic experimental durations. Although direct MD cannot fully bridge this gap, uMLIPs make it possible to extend finite-temperature stability assessments substantially beyond current AIMD practice.

A second limitation is that the NVT simulations used here keep the simulation cell fixed and therefore are not designed to fully capture phase transformations that require changes in lattice parameters, cell shape, or volume. Such cell-changing transformations may be suppressed even when they are thermodynamically favorable. Capturing these processes would require variable-cell approaches, such as NPT or Parrinello–Rahman dynamics,\cite{parrinello1980crystal} which remain challenging for high-throughput uMLIP-driven MD simulations when pressure predictions are unreliable at the relevant temperatures.\cite{li2026melt} Therefore, the MD results reported here should be interpreted as a stringent but not definitive test of finite-temperature structural robustness, complementary to harmonic phonon calculations, while longer-timescale and variable-cell simulations remain important directions for future validation.

\section{Conclusion}
In this work, we developed a hierarchical screening framework to improve the reliability of computational stability predictions and the identification of experimentally synthesizable materials. Starting from 894 materials previously reported as stable in the CAMD dataset, we curated 603 unique candidates and reassessed their thermodynamic stability using reconstructed PBE phase diagrams, of which only 298 remained stable. Dynamical-stability evaluation using uMLIP-enabled harmonic phonon calculations and nanosecond-scale MD simulations identified 166 candidates satisfying both criteria, and SCAN-based thermodynamic refinement further reduced this set to 109 stable materials. By combining decomposition enthalpy with chemical-space completeness, we ultimately prioritized 25 candidates as high-confidence targets for experimental synthesis.

Beyond identifying these candidates, this study highlights three principles for computational materials discovery. First, predicted stability should be interpreted in the context of chemical-space completeness, because incomplete competing-phase sets can overestimate synthesizability. Second, harmonic phonon calculations and finite-temperature MD simulations probe distinct instability mechanisms and should be combined rather than used interchangeably. Third, uMLIPs make it practical to extend dynamical-stability assessments far beyond the short AIMD timescales common in the literature, while allowing higher-fidelity thermodynamic methods such as SCAN to be applied selectively to the most promising candidates. Overall, this work establishes a rigorous and scalable protocol for narrowing computational predictions to candidates with stronger prospects for experimental realization.

\section{Methods}
\subsection{Density functional theory calculations}
All density functional theory (DFT) calculations were performed using the Vienna \textit{ab initio} Simulation Package (VASP) with the projector augmented-wave method.\cite{kresse1996efficiency,blochl1994projector} Structural relaxations and total energies were calculated using the Perdew--Burke--Ernzerhof (PBE) generalized-gradient approximation (GGA) functional.\cite{perdew1996generalized} Hubbard $U$ corrections\cite{wang2006oxidation,jain2011formation} were applied to transition metals in oxides, with the corrections and all other input parameters chosen consistently with those used by the Materials Project.

For calculations using the Strongly Constrained and Appropriately Normed (SCAN) meta-GGA functional,\cite{sun2015strongly} structural relaxations and total energies were computed with Hubbard $U$ corrections from Wang \textit{et al.}\cite{wang2020predicting} applied to transition metals in oxides. For all calculations, the plane-wave kinetic-energy cutoff was set to 520 eV, and the Brillouin zone was sampled using $k$-point meshes with a reciprocal-space density of 100 \r{A}$^{-3}$. Total energies and ionic forces were converged to within $10^{-5}$ eV and 0.02 eV \r{A}$^{-1}$, respectively.

\subsection{Thermodynamic stability analysis}
The thermodynamic stability of each material was evaluated using the energy above the convex hull, $E_{\rm hull}$.\cite{ong2008li} $E_{\rm hull}$ is defined as the energy difference between a compound and the corresponding linear combination of thermodynamically stable phases at the same composition in the phase diagram. For the PBE phase diagram, formation energies of competing phases were obtained from the Materials Project database (v2022.10.28),\cite{jain2013commentary} except for compounds taken from the CAMD dataset. For the SCAN phase diagram, crystal structures of all relevant compounds in the specified chemical space were retrieved from the Materials Project database (v2022.10.28),\cite{jain2013commentary} and their energies were recalculated using the SCAN functional to maintain consistency in the computational parameters. Candidate materials were evaluated using the same SCAN settings.

\subsection{Dynamical stability analysis}
Dynamical stability was evaluated using phonon dispersion calculations and molecular dynamics (MD) simulations performed with MACE,\cite{batatia2022mace} a universal machine-learning interatomic potential (uMLIP). MACE was employed as a computationally efficient surrogate for DFT, enabling simulations of larger systems and longer timescales than would be practical with first-principles methods. Unless otherwise stated, all dynamical stability analyses in this work were carried out using the pretrained MACE-MP-0 model.

Phonon dispersions were calculated using the small-displacement method\cite{alfe2009phon} as implemented in the Atomic Simulation Environment (ASE) package.\cite{larsen2017atomic} Unlike their DFT counterparts, phonon calculations performed with uMLIPs require larger supercells to obtain converged results (Figure~\ref{fig:bandstructure-benchmark}). Accordingly, supercells with a minimum length of 30~\AA{} along each lattice vector were constructed to ensure convergence of the dynamical matrix. Prior to the phonon calculations, all structures were optimized until the maximum residual force fell below $1 \times 10^{-3}$~eV~\AA$^{-1}$. A displacement amplitude of 0.05~\AA{} was used in all force-constant calculations.

MD simulations were performed in the NVT ensemble using an Andersen thermostat\cite{frenkel2023understanding} as implemented in ASE.\cite{larsen2017atomic} Supercells with minimum dimensions of 10~\r{A} in each direction were constructed for these simulations. The initial structures were first fully optimized with a force convergence criterion of 0.02 eV \r{A}$^{-1}$. MD simulations were then performed for 1~ns at an elevated temperature of 1000 K. For structures that melted as a result of exceeding the material's melting point, the simulations were instead carried out at 500 K, and we confirmed that no structures melted at this lower temperature. The NVT ensemble and elevated temperatures were used to accelerate possible structural transformations, which may otherwise occur on timescales longer than those accessible to MD. A timestep of 1~fs was used throughout the simulations. The atomic configurations were recorded every 50 fs for subsequent analysis. For each trajectory, the time-averaged potential energy was converted to an MD-averaged formation potential energy per formula unit,
\[
    \left\langle \Delta E_{\mathrm{f}} \right\rangle
    =
    \frac{
    \left\langle E_{\mathrm{pot}} \right\rangle
    - \sum_i n_i e_i^{\mathrm{ref}}
    }{
    N_{\mathrm{f.u.}}
    }
\]
where \(\left\langle E_{\mathrm{pot}} \right\rangle\) is the
trajectory-averaged potential energy of the supercell evaluated by MACE,
\(n_i\) is the number of atoms of element \(i\) in the supercell,
\(e_i^{\mathrm{ref}}\) is the per-atom energy of the most stable
\(0~\mathrm{K}\) elemental reference phase of element \(i\) calculated with
MACE, and \(N_{\mathrm{f.u.}}\) is the number of formula units in the
supercell.

Structural transitions in the MD trajectories were identified by detecting change points in the formation-energy evolution using the pruned exact linear time (PELT) with an L1-norm cost function, as implemented in the ruptures package.\cite{TRUONG2020107299} These change points were treated as indicators of potential phase transitions or structural instabilities during the simulations. To characterize the structural changes associated with each transition, radial distribution functions (RDFs) were computed for configurations before and after each change point. Specifically, RDFs for all possible atomic pairs were calculated from 10 ps trajectory segments immediately preceding and following each detected change point. The magnitude of the structural change was quantified by integrating the absolute difference between the time-averaged RDFs of the two consecutive segments:
\[
\Delta_{\rm RDF} = \int_{0}^{r_{\rm cut}}
\left|
\langle g(r)\rangle_{\rm before}
-
\langle g(r)\rangle_{\rm after}
\right| dr,
\]
where $\langle g(r)\rangle_{\rm before}$ and $\langle g(r)\rangle_{\rm after}$ are the time-averaged RDFs from the trajectory segments before and after the change point, respectively. The cutoff radius, $r_{\rm cut}$, was set to 10 \r{A}.

\subsection{Materials selection}
The materials considered for phase stability prediction were obtained from the Computational Autonomy for Materials Discovery (CAMD) dataset of \citet{ye2022novel}. The dataset comprises 894 novel materials with PBE-calculated energies above the convex hull, $E^{\rm PBE}_{\rm hull}$, within 1 meV/atom, which were reported as newly identified stable materials. In this work, duplicate crystal structures were identified using the structure-matching algorithm implemented in pymatgen with default parameters.\cite{ong2013python}

\begin{acknowledgement}
This work was supported by the City University of Hong Kong Start-up Grant (No. 9020004). Some calculations were performed on the CityU Burgundy computational facilities, managed and provided by the Computing Services Centre at the City University of Hong Kong.

\end{acknowledgement}

\begin{suppinfo}
Duplicate structures in the CAMD dataset and the entries retained after deduplication; structures matched to existing Materials Project or ICSD records; the 298 thermodynamically stable materials on the complete PBE phase diagrams; the 166 dynamically stable materials, with $E^{\mathrm{PBE}}_{\mathrm{hull}}$ and $E^{\mathrm{SCAN}}_{\mathrm{hull}}$; the synthesizability ranking of the 109 stable materials, with decomposition enthalpy $\Delta H_{\mathrm{d}}$ and composite score $S$; and convergence of the MACE-computed phonon spectra of Si with supercell size.
\end{suppinfo}

\section{Data availability}
All the data supporting this study are included in the article and its supplementary information.

\clearpage

\bibliography{ref}
\end{document}


\clearpage

\newgeometry{margin=1.4cm}
\begingroup
\scriptsize
\setlength\tabcolsep{3pt}
\renewcommand{\arraystretch}{1.15}
\setlength{\LTcapwidth}{\textwidth}
\begin{longtable}{@{}l ll cl @{\hspace{1em}} l ll cl@{}}
\caption{Seventy-two duplicate structures in the CAMD dataset. For each material, the table lists the chemical formula, space groups determined in this work and reported by CAMD, $E^{\mathrm{PBE}}_{\mathrm{hull}}$ (meV/atom), and the CAMD database ID. Bold entries indicate the structure retained after deduplication (the one with the lowest $E^{\mathrm{PBE}}_{\mathrm{hull}}$).}
\label{tab:duplicates} \\
\toprule
 & \multicolumn{2}{c}{Space group} & & &
 & \multicolumn{2}{c}{Space group} & & \\
\cmidrule(lr){2-3} \cmidrule(lr){7-8}
Material & This work & CAMD & $E^{\mathrm{PBE}}_{\mathrm{hull}}$ & Database ID &
Material & This work & CAMD & $E^{\mathrm{PBE}}_{\mathrm{hull}}$ & Database ID \\
\midrule
\endfirsthead
\multicolumn{10}{c}{{\bfseries \tablename\ \thetable{} -- continued from previous page}} \\
\toprule
 & \multicolumn{2}{c}{Space group} & & &
 & \multicolumn{2}{c}{Space group} & & \\
\cmidrule(lr){2-3} \cmidrule(lr){7-8}
Material & This work & CAMD & $E^{\mathrm{PBE}}_{\mathrm{hull}}$ & Database ID &
Material & This work & CAMD & $E^{\mathrm{PBE}}_{\mathrm{hull}}$ & Database ID \\
\midrule
\endhead
\midrule \multicolumn{10}{r}{{\footnotesize Continued on next page}} \\
\endfoot
\bottomrule
\endlastfoot
      \multirow{2}{*}{\ce{BaAgP}} & $P6_{3}/mmc$ (194) & $Pnma$ (62) & 0.7 & \textbf{camd-40815} & \multirow{2}{*}{\ce{MgB2Ir3}} & $P6/mmm$ (191) & $Cmmm$ (65) & 0.0 & \textbf{camd-10521} \\*
       & $P6_{3}/mmc$ (194) & $Cmcm$ (63) & 3.6 & camd-72895 &  & $P6/mmm$ (191) & $P6/mmm$ (191) & 0.0 & camd-287 \\
      \multirow{3}{*}{\ce{BaBr2}} & $Fm\bar{3}m$ (225) & $Fm\bar{3}m$ (225) & 0.0 & \textbf{camd-14412} & \multirow{3}{*}{\ce{MgB2Rh3}} & $P6/mmm$ (191) & $P6/mmm$ (191) & 10.0 & \textbf{camd-81729} \\*
       & $Fm\bar{3}m$ (225) & $P\bar{4}3m$ (215) & 0.0 & camd-35034 &  & $P6/mmm$ (191) & $Cmmm$ (65) & 10.2 & camd-71935 \\*
       & $Fm\bar{3}m$ (225) & $P4/mmm$ (123) & 0.1 & camd-11010 &  & $P6/mmm$ (191) & $P6_{3}/mcm$ (193) & 13.1 & camd-65711 \\
      \multirow{2}{*}{\ce{BaCl2}} & $Fm\bar{3}m$ (225) & $Pa\bar{3}$ (205) & 0.0 & \textbf{camd-28557} & \multirow{2}{*}{\ce{MgI2}} & $Cmce$ (64) & $Cmce$ (64) & 0.0 & \textbf{camd-26956} \\*
       & $Fm\bar{3}m$ (225) & $P\bar{4}2m$ (111) & 0.1 & camd-7388 &  & $C2/m$ (12) & $P2_{1}/m$ (11) & 0.4 & camd-15586 \\
      \multirow{5}{*}{\ce{BaCuP}} & $P6_{3}/mmc$ (194) & $Pnma$ (62) & 0.0 & \textbf{camd-167164} & \multirow{2}{*}{\ce{MgRuO4}} & $P2/c$ (13) & $P2/c$ (13) & 20.5 & \textbf{camd-78292} \\*
       & $P6_{3}/mmc$ (194) & $Cmcm$ (63) & 0.4 & camd-75925 &  & $P2/c$ (13) & $P\bar{1}$ (2) & 24.9 & camd-40616 \\*
       & $P6_{3}/mmc$ (194) & $Cmc2_{1}$ (36) & 0.9 & camd-72849 & \multirow{2}{*}{\ce{MgTcO4}} & $P2/c$ (13) & $P2/c$ (13) & 0.0 & \textbf{camd-49669} \\*
       & $P\bar{6}m2$ (187) & $Amm2$ (38) & 1.7 & \textbf{camd-173512} &  & $P2/c$ (13) & $P\bar{1}$ (2) & 0.2 & camd-46158 \\*
       & $P\bar{6}m2$ (187) & $P\bar{6}m2$ (187) & 2.7 & camd-75866 & \multirow{2}{*}{\ce{MgVO2}} & $R\bar{3}m$ (166) & $C2/m$ (12) & 76.7 & \textbf{camd-49161} \\*
      \multirow{4}{*}{\ce{BaF2}} & $Fm\bar{3}m$ (225) & $P2_{1}3$ (198) & 0.0 & \textbf{camd-28773} &  & $R\bar{3}m$ (166) & $C2/c$ (15) & 76.8 & camd-179036 \\*
       & $Fm\bar{3}m$ (225) & $Pa\bar{3}$ (205) & 0.0 & camd-1017 & \multirow{2}{*}{\ce{MnF3}} & $Pm\bar{3}m$ (221) & $P4/nmm$ (129) & 21.1 & \textbf{camd-25516} \\*
       & $Fm\bar{3}m$ (225) & $P4/nbm$ (125) & 0.0 & camd-3375 &  & $Pm\bar{3}m$ (221) & $Pm\bar{3}m$ (221) & 23.1 & camd-16304 \\*
       & $Fm\bar{3}m$ (225) & $Pnnm$ (58) & 0.3 & camd-6208 & \multirow{2}{*}{\ce{Na(Ni2P)2}} & $P4_{2}/mnm$ (136) & $Pnnm$ (58) & 0.0 & \textbf{camd-107223} \\*
      \multirow{3}{*}{\ce{BaHfO3}} & $Pm\bar{3}m$ (221) & $Pnma$ (62) & 0.0 & \textbf{camd-7856} &  & $P4_{2}/mnm$ (136) & $P4_{2}/mnm$ (136) & 0.4 & camd-94875 \\*
       & $Pm\bar{3}m$ (221) & $Imma$ (74) & 0.6 & camd-7849 & \multirow{2}{*}{\ce{NaCl}} & $Fm\bar{3}m$ (225) & $I4_{1}/a$ (88) & 0.0 & \textbf{camd-33814} \\*
       & $Pm\bar{3}m$ (221) & $R\bar{3}c$ (167) & 0.7 & camd-31831 &  & $Fm\bar{3}m$ (225) & $P31m$ (157) & 0.0 & camd-31018 \\
      \multirow{3}{*}{\ce{BaI2}} & $Fm\bar{3}m$ (225) & $P\bar{4}3m$ (215) & 0.0 & \textbf{camd-20756} & \multirow{2}{*}{\ce{NaWO3}} & $Cmcm$ (63) & $Cmcm$ (63) & 4.2 & \textbf{camd-122014} \\*
       & $Fm\bar{3}m$ (225) & $Fm\bar{3}m$ (225) & 0.0 & camd-5730 &  & $Pnma$ (62) & $Pnma$ (62) & 22.7 & camd-52250 \\*
       & $Fm\bar{3}m$ (225) & $I4/mmm$ (139) & 0.1 & camd-6393 & \multirow{2}{*}{\ce{NaZnP}} & $P6_{3}/mmc$ (194) & $Cc$ (9) & 0.0 & \textbf{camd-9134} \\*
      \multirow{2}{*}{\ce{BeF2}} & $P6_{3}/mmc$ (194) & $P6_{3}/mmc$ (194) & 2.4 & \textbf{camd-13618} &  & $P6_{3}/mmc$ (194) & $Cmc2_{1}$ (36) & 0.4 & camd-24442 \\*
       & $Cmc2_{1}$ (36) & $Cmc2_{1}$ (36) & 2.7 & camd-32808 & \multirow{2}{*}{\ce{NbBr5}} & $P6_{3}/mmc$ (194) & $P6_{3}/mmc$ (194) & 9.4 & \textbf{camd-24148} \\*
      \multirow{2}{*}{\ce{Ca(CoP)2}} & $I4/mmm$ (139) & $C2/m$ (12) & 0.0 & \textbf{camd-39114} &  & $Cmcm$ (63) & $Cmcm$ (63) & 9.5 & camd-222 \\*
       & $I4/mmm$ (139) & $Immm$ (71) & 1.9 & camd-61166 & \multirow{2}{*}{\ce{RbAlO2}} & $Pna2_{1}$ (33) & $Pna2_{1}$ (33) & 0.0 & \textbf{camd-84589} \\*
      \multirow{2}{*}{\ce{Ca(ZnP)2}} & $P\bar{3}m1$ (164) & $P2_{1}/m$ (11) & 0.0 & \textbf{camd-49022} &  & $P4_{1}2_{1}2$ (92) & $P4_{1}2_{1}2$ (92) & 0.6 & camd-70630 \\*
       & $P\bar{3}m1$ (164) & $P\bar{3}$ (147) & 3.2 & camd-150708 & \multirow{2}{*}{\ce{SbBr5}} & $P6_{3}/mmc$ (194) & $Cmcm$ (63) & 51.4 & \textbf{camd-9257} \\*
      \multirow{3}{*}{\ce{Ca3BiP}} & $Pnma$ (62) & $Pnma$ (62) & 0.0 & \textbf{camd-12078} &  & $P6_{3}/mmc$ (194) & $P6_{3}/mmc$ (194) & 51.7 & camd-6394 \\*
       & $Pnma$ (62) & $P2_{1}/c$ (14) & 0.0 & camd-15021 & \multirow{2}{*}{\ce{ScSb}} & $P6_{3}/mmc$ (194) & $P6_{3}/mmc$ (194) & 8.7 & \textbf{camd-42127} \\*
       & $Pnma$ (62) & $P\bar{1}$ (2) & 0.1 & camd-26873 &  & $P6_{3}/mmc$ (194) & $P6_{3}mc$ (186) & 9.5 & camd-87503 \\
      \multirow{2}{*}{\ce{Ca4GaP}} & $P4/mmm$ (123) & $P4/mmm$ (123) & 0.0 & \textbf{camd-62024} & \multirow{2}{*}{\ce{SiCl4}} & $I\bar{4}3m$ (217) & $I\bar{4}2m$ (121) & 8.8 & \textbf{camd-1083} \\*
       & $P4/mmm$ (123) & $Pmmm$ (47) & 0.1 & camd-58422 &  & $I\bar{4}3m$ (217) & $I\bar{4}3m$ (217) & 9.2 & camd-1079 \\
      \multirow{2}{*}{\ce{CaF2}} & $Fm\bar{3}m$ (225) & $Pa\bar{3}$ (205) & 0.0 & \textbf{camd-27795} & \multirow{2}{*}{\ce{Sr(IrO3)2}} & $Imma$ (74) & $Imma$ (74) & 0.0 & \textbf{camd-35983} \\*
       & $Fm\bar{3}m$ (225) & $P4/nbm$ (125) & 0.1 & camd-15704 &  & $Fd\bar{3}m$ (227) & $Fd\bar{3}m$ (227) & 0.1 & camd-27521 \\
      \multirow{2}{*}{\ce{CaHfP2}} & $R\bar{3}m$ (166) & $R\bar{3}m$ (166) & 0.0 & \textbf{camd-56240} & \multirow{2}{*}{\ce{Sr(NiP2)2}} & $I4/mmm$ (139) & $Immm$ (71) & 0.6 & \textbf{camd-171057} \\*
       & $R\bar{3}m$ (166) & $C2/m$ (12) & 1.1 & camd-52276 &  & $I4/mmm$ (139) & $I4/mmm$ (139) & 4.1 & camd-149751 \\
      \multirow{2}{*}{\ce{CaZrP2}} & $R\bar{3}m$ (166) & $R\bar{3}m$ (166) & 0.0 & \textbf{camd-174805} & \multirow{3}{*}{\ce{Sr(ZnP)2}} & $P\bar{3}m1$ (164) & $P2/m$ (10) & 0.0 & \textbf{camd-53893} \\*
       & $R\bar{3}m$ (166) & $C2/m$ (12) & 4.9 & camd-54716 &  & $P\bar{3}m1$ (164) & $P\bar{3}$ (147) & 3.3 & camd-76880 \\*
      \multirow{2}{*}{\ce{CoSn2}} & $Fm\bar{3}m$ (225) & $P2_{1}3$ (198) & 3.4 & \textbf{camd-32645} &  & $P\bar{3}m1$ (164) & $C2/m$ (12) & 4.7 & camd-48052 \\*
       & $Fm\bar{3}m$ (225) & $Fm\bar{3}m$ (225) & 6.9 & camd-21567 & \multirow{3}{*}{\ce{Sr2ZrP2}} & $P\bar{3}m1$ (164) & $C2/m$ (12) & 107.6 & \textbf{camd-43637} \\*
      \multirow{3}{*}{\ce{CsBr}} & $Fm\bar{3}m$ (225) & $Immm$ (71) & 0.2 & \textbf{camd-36213} &  & $P\bar{3}m1$ (164) & $P\bar{3}m1$ (164) & 107.7 & camd-127538 \\*
       & $Immm$ (71) & $Imm2$ (44) & 0.5 & camd-9375 &  & $C2/m$ (12) & $Cm$ (8) & 108.1 & camd-69092 \\*
       & $P2_{1}3$ (198) & $P2_{1}3$ (198) & 1.7 & camd-4555 & \multirow{2}{*}{\ce{SrAgP}} & $P6_{3}/mmc$ (194) & $Cmcm$ (63) & 4.2 & \textbf{camd-79396} \\*
      \multirow{4}{*}{\ce{CsF}} & $Immm$ (71) & $Immm$ (71) & 0.4 & \textbf{camd-8883} &  & $P6_{3}/mmc$ (194) & $P6_{3}mc$ (186) & 4.3 & camd-42862 \\*
       & $Fm\bar{3}m$ (225) & $C2/m$ (12) & 0.4 & camd-11796 & \multirow{2}{*}{\ce{SrCuP}} & $P6_{3}/mmc$ (194) & $Pnma$ (62) & 0.2 & \textbf{camd-55079} \\*
       & $Fm\bar{3}m$ (225) & $I4/mmm$ (139) & 0.4 & camd-13063 &  & $P6_{3}/mmc$ (194) & $Cmc2_{1}$ (36) & 1.4 & camd-82676 \\*
       & $Fm\bar{3}m$ (225) & $P2_{1}3$ (198) & 0.5 & camd-11804 & \multirow{2}{*}{\ce{SrF2}} & $Fm\bar{3}m$ (225) & $P2_{1}3$ (198) & 0.1 & \textbf{camd-36029} \\*
      \multirow{3}{*}{\ce{FeN}} & $F\bar{4}3m$ (216) & $I\bar{4}m2$ (119) & 0.1 & \textbf{camd-7258} &  & $Fm\bar{3}m$ (225) & $P\bar{4}2m$ (111) & 0.1 & camd-22965 \\*
       & $F\bar{4}3m$ (216) & $Pmn2_{1}$ (31) & 1.6 & camd-10475 & \multirow{2}{*}{\ce{SrTiO3}} & $Cmcm$ (63) & $Cmcm$ (63) & 0.6 & \textbf{camd-77955} \\*
       & $F\bar{4}3m$ (216) & $P3_{1}$ (144) & 2.5 & camd-33824 &  & $Pnma$ (62) & $Pnma$ (62) & 0.8 & camd-52046 \\
      \multirow{2}{*}{\ce{K3AuO}} & $Pm\bar{3}m$ (221) & $Pmmm$ (47) & 0.0 & \textbf{camd-163229} & \multirow{2}{*}{\ce{TlCl3}} & $P\bar{3}1c$ (163) & $P\bar{3}1c$ (163) & 0.6 & \textbf{camd-37503} \\*
       & $Pm\bar{3}m$ (221) & $Amm2$ (38) & 0.0 & camd-53088 &  & $C2/c$ (15) & $C2/c$ (15) & 3.2 & camd-14696 \\
      \multirow{3}{*}{\ce{KBr}} & $Fm\bar{3}m$ (225) & $Immm$ (71) & 0.0 & \textbf{camd-10071} & \multirow{2}{*}{\ce{VBr5}} & $P6_{3}/mmc$ (194) & $P6_{3}/mmc$ (194) & 28.6 & \textbf{camd-36016} \\*
       & $Fm\bar{3}m$ (225) & $Fmmm$ (69) & 0.0 & camd-8847 &  & $Cmcm$ (63) & $Cmcm$ (63) & 29.0 & camd-16802 \\*
       & $Fm\bar{3}m$ (225) & $P3_{2}21$ (154) & 0.1 & camd-29910 & \multirow{2}{*}{\ce{VCl5}} & $P6_{3}/mmc$ (194) & $Cmcm$ (63) & 0.0 & \textbf{camd-36465} \\*
      \multirow{2}{*}{\ce{LiTi2N3}} & $Immm$ (71) & $Pmmn$ (59) & 0.0 & \textbf{camd-15097} &  & $P6_{3}/mmc$ (194) & $P6_{3}/mmc$ (194) & 0.3 & camd-28701 \\*
       & $Immm$ (71) & $Immm$ (71) & 2.4 & camd-17063 & \multirow{2}{*}{\ce{VS2}} & $Fmm2$ (42) & $Fmm2$ (42) & 0.0 & \textbf{camd-32349} \\*
      \multirow{2}{*}{\ce{Mg(ZnP)2}} & $P\bar{3}m1$ (164) & $P2/m$ (10) & 0.0 & \textbf{camd-56520} &  & $Fmm2$ (42) & $Amm2$ (38) & 2.9 & camd-4254 \\*
       & $P\bar{3}m1$ (164) & $P\bar{3}m1$ (164) & 6.0 & camd-41614 & \multirow{2}{*}{\ce{YAs}} & $Fm\bar{3}m$ (225) & $Immm$ (71) & 0.0 & \textbf{camd-74230} \\*
      \multirow{2}{*}{\ce{Mg2Co3P}} & $P6_{3}/mmc$ (194) & $Cmcm$ (63) & 8.1 & \textbf{camd-179994} &  & $Fm\bar{3}m$ (225) & $I4/mmm$ (139) & 0.0 & camd-39154 \\*
       & $P6_{3}/mmc$ (194) & $P6_{3}/mmc$ (194) & 8.2 & camd-117214 & \multirow{2}{*}{\ce{YSb}} & $Fm\bar{3}m$ (225) & $I4/mmm$ (139) & 0.0 & \textbf{camd-40988} \\*
      \multirow{2}{*}{\ce{Mg2TiO4}} & $Fd\bar{3}m$ (227) & $Fd\bar{3}m$ (227) & 0.0 & \textbf{camd-49983} &  & $Fm\bar{3}m$ (225) & $Immm$ (71) & 0.0 & camd-68300 \\*
       & $Fd\bar{3}m$ (227) & $Imma$ (74) & 0.0 & camd-79378 & \multirow{2}{*}{\ce{Zn(RhSe2)2}} & $Fd\bar{3}m$ (227) & $Imma$ (74) & 16.8 & \textbf{camd-61323} \\*
      & & & & &  & $Fd\bar{3}m$ (227) & $Fd\bar{3}m$ (227) & 16.8 & camd-39432 \\
\end{longtable}
\endgroup
\restoregeometry

\clearpage
\newgeometry{margin=1.8cm} 
\begingroup
\footnotesize
\setlength\tabcolsep{4pt}
\renewcommand{\arraystretch}{1.15}
\setlength{\LTcapwidth}{\textwidth}

\begin{longtable}{@{}llll @{\hspace{2.2em}} llll@{}}
\caption{Seventy-five structures in the CAMD dataset matched to existing records in the Materials Project (MP) and the Inorganic Crystal Structure Database (ICSD). Of these, 59 matched ICSD records and 16 matched MP records. For each material, the table lists the chemical formula, space group, the CAMD database ID, and the corresponding MP or ICSD ID.}
\label{tab:matched_ss} \\
\toprule
& & \multicolumn{2}{c}{Database ID} & & & \multicolumn{2}{c}{Database ID} \\
\cmidrule(lr){3-4} \cmidrule(lr){7-8}
Material & Space group & CAMD & MP\,/\,ICSD &
Material & Space group & CAMD & MP\,/\,ICSD \\
\midrule
\endfirsthead

\multicolumn{8}{c}{{\bfseries \tablename\ \thetable{} -- continued from previous page}} \\
\toprule
& & \multicolumn{2}{c}{Database ID} & & & \multicolumn{2}{c}{Database ID} \\
\cmidrule(lr){3-4} \cmidrule(lr){7-8}
Material & Space group & CAMD & MP\,/\,ICSD &
Material & Space group & CAMD & MP\,/\,ICSD \\
\midrule
\endhead

\midrule \multicolumn{8}{r}{{\footnotesize Continued on next page}} \\
\endfoot

\bottomrule
\endlastfoot
      \ce{Ba(FeAs)2} & $I4/mmm$ (139) & camd-7517 & icsd-166018 & \ce{MgCl2} & $P\bar{1}$ (2) & camd-9293 & icsd-26157 \\
      \ce{Ba2RuO4} & $I4/mmm$ (139) & camd-28544 & mp-1025237 & \ce{MgSbPd2} & $Fm\bar{3}m$ (225) & camd-111755 & mp-977579 \\
      \ce{BaAgP} & $P6_{3}/mmc$ (194) & camd-40815 & icsd-41706 & \ce{MgSbPt} & $F\bar{4}3m$ (216) & camd-7 & icsd-44826 \\
      \ce{BaCl2} & $Fm\bar{3}m$ (225) & camd-28557 & icsd-2191 & \ce{MgV2O4} & $I4_{1}/amd$ (141) & camd-167964 & icsd-56283 \\
      \ce{BaCuP} & $P6_{3}/mmc$ (194) & camd-167164 & icsd-52684 & \ce{MnAgO2} & $Cmcm$ (63) & camd-22579 & mp-997162 \\
      \ce{BaF2} & $Fm\bar{3}m$ (225) & camd-28773 & icsd-41649 & \ce{Na2MoO4} & $Fd\bar{3}m$ (227) & camd-75541 & icsd-44523 \\
      \ce{BaHfO3} & $Pm\bar{3}m$ (221) & camd-7856 & mp-998552 & \ce{Na2WO4} & $Fd\bar{3}m$ (227) & camd-50347 & icsd-2133 \\
      \ce{BaMnO3} & $C2/m$ (12) & camd-11540 & icsd-89995 & \ce{Na4CoO4} & $P\bar{1}$ (2) & camd-50520 & mp-27224 \\
      \ce{BaTiO3} & $Amm2$ (38) & camd-22655 & icsd-31155 & \ce{NaBr} & $Fm\bar{3}m$ (225) & camd-8152 & icsd-18013 \\
      \ce{Ca(CoP)2} & $I4/mmm$ (139) & camd-39114 & icsd-10462 & \ce{NaCl} & $Fm\bar{3}m$ (225) & camd-33814 & icsd-18189 \\
      \ce{Ca(Cu2P)2} & $P4_{2}/mnm$ (136) & camd-58477 & icsd-62553 & \ce{NaF} & $Fm\bar{3}m$ (225) & camd-10 & icsd-29128 \\
      \ce{Ca(ZnP)2} & $P\bar{3}m1$ (164) & camd-49022 & icsd-100062 & \ce{NaI} & $Fm\bar{3}m$ (225) & camd-6269 & icsd-44279 \\
      \ce{CaAgP} & $P\bar{6}2m$ (189) & camd-61244 & icsd-10016 & \ce{Ni3S4} & $Fd\bar{3}m$ (227) & camd-5303 & icsd-36271 \\
      \ce{CaF2} & $Fm\bar{3}m$ (225) & camd-27795 & icsd-28730 & \ce{NiS2} & $Pa\bar{3}$ (205) & camd-1887 & icsd-40328 \\
      \ce{CaIn2O4} & $Fd\bar{3}m$ (227) & camd-50214 & mp-22766 & \ce{PdCl2} & $R\bar{3}m$ (166) & camd-2638 & icsd-404624 \\
      \ce{CaTcO3} & $Pnma$ (62) & camd-44008 & icsd-261341 & \ce{RbBr} & $Fm\bar{3}m$ (225) & camd-19807 & icsd-18017 \\
      \ce{CaVO3} & $Pnma$ (62) & camd-22320 & mp-1435359 & \ce{SbCl5} & $Cmcm$ (63) & camd-3526 & icsd-26589 \\
      \ce{CaZrO3} & $R3c$ (161) & camd-19322 & mp-776024 & \ce{ScAs} & $Fm\bar{3}m$ (225) & camd-59300 & icsd-1331 \\
      \ce{CdBr2} & $C2/m$ (12) & camd-31851 & icsd-52367 & \ce{ScF3} & $P321$ (150) & camd-15095 & icsd-77071 \\
      \ce{CsBr} & $Fm\bar{3}m$ (225) & camd-36213 & icsd-61516 & \ce{ScTe} & $P6_{3}/mmc$ (194) & camd-47451 & icsd-43593 \\
      \ce{CsF} & $Immm$ (71) & camd-8883 & icsd-44288 & \ce{Si} & $Fd\bar{3}m$ (227) & camd-40283 & icsd-29287 \\
      \ce{CuSbRh2} & $I4/mmm$ (139) & camd-96082 & mp-867753 & \ce{SiF4} & $I\bar{4}3m$ (217) & camd-24042 & icsd-14122 \\
      \ce{FeB} & $Cmcm$ (63) & camd-35505 & icsd-613898 & \ce{Sr(FeAs)2} & $I4/mmm$ (139) & camd-28554 & icsd-163208 \\
      \ce{FeN} & $F\bar{4}3m$ (216) & camd-7258 & icsd-41258 & \ce{Sr(InP)2} & $P6_{3}/mmc$ (194) & camd-71220 & icsd-260563 \\
      \ce{HgF2} & $P3_{2}21$ (154) & camd-36556 & icsd-33614 & \ce{Sr(ZnP)2} & $P\bar{3}m1$ (164) & camd-53893 & icsd-30911 \\
      \ce{K(FeP)2} & $C2/m$ (12) & camd-23591 & mp-1206409 & \ce{Sr2VO4} & $I4/mmm$ (139) & camd-48785 & icsd-69000 \\
      \ce{K3AuO} & $Pm\bar{3}m$ (221) & camd-163229 & icsd-79086 & \ce{Sr3WO6} & $P2_{1}/c$ (14) & camd-24177 & mp-755326 \\
      \ce{KBr} & $Fm\bar{3}m$ (225) & camd-10071 & icsd-18015 & \ce{SrAgP} & $P6_{3}/mmc$ (194) & camd-79396 & icsd-52596 \\
      \ce{KF} & $Fm\bar{3}m$ (225) & camd-10674 & icsd-52241 & \ce{SrCl2} & $Fm\bar{3}m$ (225) & camd-30382 & icsd-18011 \\
      \ce{KNbO3} & $Amm2$ (38) & camd-70418 & icsd-9534 & \ce{SrCuP} & $P6_{3}/mmc$ (194) & camd-55079 & icsd-53323 \\
      \ce{Li(MnS2)2} & $Fd\bar{3}m$ (227) & camd-57175 & mp-1045401 & \ce{SrF2} & $Fm\bar{3}m$ (225) & camd-36029 & icsd-40414 \\
      \ce{Li2MnO2} & $P\bar{3}m1$ (164) & camd-28113 & icsd-37327 & \ce{SrFeO2} & $P\bar{4}m2$ (115) & camd-69300 & icsd-173434 \\
      \ce{LiCrO2} & $Fd\bar{3}m$ (227) & camd-47767 & mp-756362 & \ce{SrTiO3} & $R\bar{3}c$ (167) & camd-163955 & icsd-182247 \\
      \ce{LiF} & $P6_{3}/mmc$ (194) & camd-17090 & mp-1185301 & \ce{TlCl} & $Fm\bar{3}m$ (225) & camd-25103 & icsd-61518 \\
      \ce{LiInO2} & $I4_{1}/amd$ (141) & camd-112862 & mp-1222354 & \ce{VO} & $Fm\bar{3}m$ (225) & camd-6207 & icsd-28681 \\
      \ce{LiMn2O4} & $I4_{1}/amd$ (141) & camd-74597 & icsd-40485 & \ce{YAs} & $Fm\bar{3}m$ (225) & camd-74230 & icsd-44087 \\
      \ce{LiTiO2} & $I4_{1}/amd$ (141) & camd-91641 & icsd-164158 & \ce{YSb} & $Fm\bar{3}m$ (225) & camd-40988 & icsd-43632 \\
      \ce{LiV2O4} & $Imma$ (74) & camd-154345 & mp-1288227 &  &  &  &  \\
\end{longtable}
\endgroup
\restoregeometry

\clearpage
\newgeometry{margin=1.8cm}            
\begingroup
\footnotesize
\setlength\tabcolsep{4pt}
\renewcommand{\arraystretch}{1.15}
\setlength{\LTcapwidth}{\textwidth}

\begin{longtable}{@{}llcl @{\hspace{2.2em}} llcl@{}}
\caption{The 298 thermodynamically stable materials identified from the PBE-based thermodynamic stability analysis. For each material, the table lists the chemical formula, space group, $E^{\mathrm{PBE}}_{\mathrm{hull}}$ (meV/atom), and the CAMD database ID.}
\label{tab:pbe_stables} \\
\toprule
Material & Space group & $E^{\mathrm{PBE}}_{\mathrm{hull}}$ & Database ID &
Material & Space group & $E^{\mathrm{PBE}}_{\mathrm{hull}}$ & Database ID \\
\midrule
\endfirsthead

\multicolumn{8}{c}{{\bfseries \tablename\ \thetable{} -- continued from previous page}} \\
\toprule
Material & Space group & $E^{\mathrm{PBE}}_{\mathrm{hull}}$ & Database ID &
Material & Space group & $E^{\mathrm{PBE}}_{\mathrm{hull}}$ & Database ID \\
\midrule
\endhead

\midrule \multicolumn{8}{r}{{\footnotesize Continued on next page}} \\
\endfoot

\bottomrule
\endlastfoot
\ce{Ag3Br4} & $Imma$ (74) & 0.0 & camd-9884 & \ce{Ag3Cl4} & $Imma$ (74) & 0.0 & camd-120 \\
\ce{AgCl} & $P6_{3}mc$ (186) & 0.0 & camd-123 & \ce{AgCl2} & $P2_{1}/c$ (14) & 0.0 & camd-3286 \\
\ce{AgF3} & $Im\bar{3}$ (204) & 0.0 & camd-36738 & \ce{Al(TcB)2} & $Cmmm$ (65) & 0.0 & camd-43082 \\
\ce{Al2Tc3B2} & $Cmmm$ (65) & 0.0 & camd-68367 & \ce{AlBPt3} & $R\bar{3}$ (148) & 0.0 & camd-52396 \\
\ce{AuCl5} & $Cmc2_{1}$ (36) & 0.0 & camd-15342 & \ce{Ba(Cu2P)2} & $Cmmm$ (65) & 0.0 & camd-44902 \\
\ce{Ba(InP)2} & $P2/m$ (10) & 0.0 & camd-126251 & \ce{Ba(IrO3)2} & $P\bar{3}1m$ (162) & 0.0 & camd-15889 \\
\ce{Ba(ZnP)2} & $Pnma$ (62) & 0.0 & camd-141216 & \ce{Ba2RuO5} & $P4/mmm$ (123) & 0.0 & camd-4628 \\
\ce{Ba2TcO4} & $I4/mmm$ (139) & 0.0 & camd-30164 & \ce{BaBr2} & $Fm\bar{3}m$ (225) & 0.0 & camd-14412 \\
\ce{BaFeO2} & $Ama2$ (40) & 0.0 & camd-33312 & \ce{BaI2} & $Fm\bar{3}m$ (225) & 0.0 & camd-20756 \\
\ce{BaIn2O4} & $P2_{1}/m$ (11) & 0.0 & camd-21023 & \ce{BaMn2O4} & $P2_{1}/m$ (11) & 0.0 & camd-4135 \\
\ce{BaMnO2} & $Pmmn$ (59) & 0.0 & camd-26647 & \ce{BaNiP} & $Pnma$ (62) & 0.0 & camd-43939 \\
\ce{BaPRh} & $P\bar{6}m2$ (187) & 0.0 & camd-53638 & \ce{BaPdO3} & $C2/m$ (12) & 0.0 & camd-21317 \\
\ce{BaPtO3} & $C2/c$ (15) & 0.0 & camd-12734 & \ce{BaRuO4} & $P1$ (1) & 0.0 & camd-21763 \\
\ce{BaTa2O6} & $Pmma$ (51) & 0.0 & camd-10816 & \ce{BaTcO4} & $Cmcm$ (63) & 0.0 & camd-7233 \\
\ce{BiCl3} & $P\bar{3}1c$ (163) & 0.0 & camd-29297 & \ce{Ca(IrO3)2} & $P\bar{3}1m$ (162) & 0.0 & camd-79291 \\
\ce{Ca(MnP)2} & $I4/mmm$ (139) & 0.0 & camd-94778 & \ce{Ca(PRh)2} & $P3_{1}21$ (154) & 0.0 & camd-63707 \\
\ce{Ca(PdO2)3} & $Cmmm$ (65) & 0.0 & camd-65605 & \ce{Ca2(PRh)3} & $Pmmn$ (59) & 0.0 & camd-95513 \\
\ce{Ca2AlP} & $P\bar{3}m1$ (164) & 0.0 & camd-83665 & \ce{Ca2Cd3O5} & $C2/m$ (12) & 0.0 & camd-66539 \\
\ce{Ca2Cu3P4} & $Ibam$ (72) & 0.0 & camd-170835 & \ce{Ca2IrO4} & $Cmmm$ (65) & 0.0 & camd-170316 \\
\ce{Ca2RhO4} & $Cmmm$ (65) & 0.0 & camd-46457 & \ce{Ca2ZnP} & $Amm2$ (38) & 0.0 & camd-92647 \\
\ce{Ca2ZnP2} & $P6_{3}/mmc$ (194) & 0.0 & camd-41223 & \ce{Ca2ZrP2} & $P\bar{3}m1$ (164) & 0.0 & camd-43507 \\
\ce{Ca3(GaP2)2} & $Pnnm$ (58) & 0.0 & camd-80913 & \ce{Ca3(Zn2P)2} & $P\bar{6}2m$ (189) & 0.0 & camd-166106 \\
\ce{Ca3Ag3P} & $Cmcm$ (63) & 0.0 & camd-67218 & \ce{Ca3AgO4} & $Pm\bar{3}m$ (221) & 0.0 & camd-40913 \\
\ce{Ca3AgP} & $Pnma$ (62) & 0.0 & camd-61403 & \ce{Ca3Al3P} & $P2_{1}/m$ (11) & 0.0 & camd-80123 \\
\ce{Ca3AlP3} & $P\bar{1}$ (2) & 0.0 & camd-43464 & \ce{Ca3BiP} & $Pnma$ (62) & 0.0 & camd-12078 \\
\ce{Ca3Cd2O5} & $C2/m$ (12) & 0.0 & camd-76036 & \ce{Ca3GaP} & $Pnma$ (62) & 0.0 & camd-73868 \\
\ce{Ca3InP} & $Pnma$ (62) & 0.0 & camd-44172 & \ce{Ca3MoO6} & $R\bar{3}$ (148) & 0.0 & camd-160867 \\
\ce{Ca3P2Rh3} & $Cmcm$ (63) & 0.0 & camd-78315 & \ce{Ca3RuO6} & $R\bar{3}$ (148) & 0.0 & camd-81624 \\
\ce{Ca3ZnP} & $Pnma$ (62) & 0.0 & camd-49451 & \ce{Ca4AlP} & $P2/c$ (13) & 0.0 & camd-50716 \\
\ce{Ca4CuP2} & $P4/mmm$ (123) & 0.0 & camd-74077 & \ce{Ca4GaP} & $P4/mmm$ (123) & 0.0 & camd-62024 \\
\ce{Ca4PRh} & $P4/mmm$ (123) & 0.0 & camd-46184 & \ce{CaBr2} & $P\bar{1}$ (2) & 0.0 & camd-20618 \\
\ce{CaCd5O6} & $C2/m$ (12) & 0.0 & camd-74563 & \ce{CaCoP} & $P4/nmm$ (129) & 0.0 & camd-53623 \\
\ce{CaFeP} & $P4/nmm$ (129) & 0.0 & camd-29355 & \ce{CaHf2P3} & $P\bar{3}m1$ (164) & 0.0 & camd-68524 \\
\ce{CaHfP2} & $R\bar{3}m$ (166) & 0.0 & camd-56240 & \ce{CaNiP} & $P\bar{6}m2$ (187) & 0.0 & camd-40483 \\
\ce{CaPIr} & $P2_{13}$ (198) & 0.0 & camd-56579 & \ce{CaPRh} & $P2_{1}/c$ (14) & 0.0 & camd-43504 \\
\ce{CaRuO4} & $Pbcn$ (60) & 0.0 & camd-47329 & \ce{CaTcO4} & $P2/c$ (13) & 0.0 & camd-128896 \\
\ce{CaZrP2} & $R\bar{3}m$ (166) & 0.0 & camd-174805 & \ce{CdI2} & $I\bar{4}3m$ (217) & 0.0 & camd-18153 \\
\ce{CoBr2} & $I\bar{4}2d$ (122) & 0.0 & camd-27843 & \ce{CoBr3} & $R\bar{3}$ (148) & 0.0 & camd-6836 \\
\ce{CoCl2} & $I\bar{4}3m$ (217) & 0.0 & camd-19238 & \ce{CoCl3} & $P\bar{3}1m$ (162) & 0.0 & camd-21981 \\
\ce{CoF4} & $I\bar{4}2m$ (121) & 0.0 & camd-17198 & \ce{CoI2} & $P3_{1}21$ (154) & 0.0 & camd-1649 \\
\ce{CoSn4} & $P4/nbm$ (125) & 0.0 & camd-20811 & \ce{CrCl4} & $P\bar{4}3m$ (215) & 0.0 & camd-24205 \\
\ce{CrPRu} & $Pnma$ (62) & 0.0 & camd-82918 & \ce{CrSbS4} & $C2/m$ (12) & 0.0 & camd-162358 \\
\ce{CsPtO3} & $Cmcm$ (63) & 0.0 & camd-37880 & \ce{Cu2SbRh3} & $P\bar{3}m1$ (164) & 0.0 & camd-85589 \\
\ce{Cu2TcSe4} & $P\bar{4}2m$ (111) & 0.0 & camd-98678 & \ce{Cu3B3Pt4} & $I\bar{4}3d$ (220) & 0.0 & camd-64822 \\
\ce{Cu3Br4} & $P\bar{4}3m$ (215) & 0.0 & camd-35151 & \ce{Cu3Cl4} & $P\bar{4}3m$ (215) & 0.0 & camd-20937 \\
\ce{Cu3I4} & $P\bar{4}3m$ (215) & 0.0 & camd-23317 & \ce{Cu3TcSe4} & $P\bar{4}3m$ (215) & 0.0 & camd-176274 \\
\ce{CuBPt2} & $Cmcm$ (63) & 0.0 & camd-108676 & \ce{CuF3} & $Im\bar{3}$ (204) & 0.0 & camd-9764 \\
\ce{CuTc3Se4} & $Cmc2_{1}$ (36) & 0.0 & camd-183475 & \ce{CuTcSb4} & $P2/m$ (10) & 0.0 & camd-108054 \\
\ce{CuTcSe4} & $P2/m$ (10) & 0.0 & camd-89283 & \ce{CuTe2Au3} & $R\bar{3}m$ (166) & 0.0 & camd-40920 \\
\ce{Fe2CoP2Ru} & $Pmc2_{1}$ (26) & 0.0 & camd-111559 & \ce{Fe2NiP2Ru} & $Pmc2_{1}$ (26) & 0.0 & camd-89058 \\
\ce{FeI3} & $Ama2$ (40) & 0.0 & camd-33136 & \ce{GaF3} & $Fd\bar{3}m$ (227) & 0.0 & camd-19356 \\
\ce{GeF3} & $Fm\bar{3}m$ (225) & 0.0 & camd-32343 & \ce{InCl3} & $C2/c$ (15) & 0.0 & camd-22548 \\
\ce{IrCl5} & $P\bar{1}$ (2) & 0.0 & camd-33422 & \ce{IrF3} & $Fd\bar{3}m$ (227) & 0.0 & camd-17173 \\
\ce{K(CuP2)2} & $I4/mmm$ (139) & 0.0 & camd-152270 & \ce{K(FeN)2} & $C2/m$ (12) & 0.0 & camd-2275 \\
\ce{K2HfO3} & $Cmcm$ (63) & 0.0 & camd-82505 & \ce{K2HfS3} & $C2/m$ (12) & 0.0 & camd-34482 \\
\ce{K3Al2P3} & $P2_{1}/m$ (11) & 0.0 & camd-49812 & \ce{K3AlP2} & $Ibam$ (72) & 0.0 & camd-69436 \\
\ce{K3AuO2} & $P4_{2}/mnm$ (136) & 0.0 & camd-53418 & \ce{K3CoO3} & $Cmcm$ (63) & 0.0 & camd-75314 \\
\ce{K3In2P3} & $P2_{1}/m$ (11) & 0.0 & camd-43890 & \ce{K3InO3} & $P\bar{1}$ (2) & 0.0 & camd-41331 \\
\ce{KAuO} & $Cmcm$ (63) & 0.0 & camd-59256 & \ce{KFeO2} & $Pna2_{1}$ (33) & 0.0 & camd-64660 \\
\ce{KFeP} & $P4/nmm$ (129) & 0.0 & camd-12435 & \ce{KPtO3} & $Fd\bar{3}m$ (227) & 0.0 & camd-47276 \\
\ce{Li(CoP)2} & $I4/mmm$ (139) & 0.0 & camd-35644 & \ce{Li(Ni2P)2} & $P4_{2}/mnm$ (136) & 0.0 & camd-267 \\
\ce{Li2RuO3} & $C2/c$ (15) & 0.0 & camd-68313 & \ce{Li2TcO4} & $I\bar{4}2d$ (122) & 0.0 & camd-171851 \\
\ce{Li3AgO3} & $P4_{2}/mnm$ (136) & 0.0 & camd-115567 & \ce{Li3RuO4} & $C2/c$ (15) & 0.0 & camd-75839 \\
\ce{Li4CdO3} & $R\bar{3}m$ (166) & 0.0 & camd-61558 & \ce{Li4MoO5} & $C2/c$ (15) & 0.0 & camd-78244 \\
\ce{Li4RuO5} & $C2/c$ (15) & 0.0 & camd-44905 & \ce{LiAgO2} & $P4_{2}/mmc$ (131) & 0.0 & camd-159085 \\
\ce{LiAlO2} & $Pna2_{1}$ (33) & 0.0 & camd-131277 & \ce{LiAuCl4} & $I4_{1}md$ (109) & 0.0 & camd-25070 \\
\ce{LiCoP} & $P4/nmm$ (129) & 0.0 & camd-3466 & \ce{LiCuO2} & $P2_{1}/c$ (14) & 0.0 & camd-66501 \\
\ce{LiRuO3} & $Pccn$ (56) & 0.0 & camd-45835 & \ce{LiTcO4} & $C2$ (5) & 0.0 & camd-38153 \\
\ce{LiTi2N3} & $Immm$ (71) & 0.0 & camd-15097 & \ce{Mg(PdO2)3} & $Cmmm$ (65) & 0.0 & camd-45298 \\
\ce{Mg(RuO3)2} & $P4_{2}/mnm$ (136) & 0.0 & camd-53651 & \ce{Mg(TcB)2} & $Cmmm$ (65) & 0.0 & camd-37677 \\
\ce{Mg(ZnP)2} & $P\bar{3}m1$ (164) & 0.0 & camd-56520 & \ce{Mg2BIr2} & $Pbam$ (55) & 0.0 & camd-22036 \\
\ce{Mg2BPt4} & $Cmmm$ (65) & 0.0 & camd-37034 & \ce{Mg2BRh2} & $Pbam$ (55) & 0.0 & camd-45654 \\
\ce{Mg2Ni3P} & $P6_{3}/mmc$ (194) & 0.0 & camd-134578 & \ce{Mg2SbRh3} & $P\bar{3}m1$ (164) & 0.0 & camd-185530 \\
\ce{Mg2TiO4} & $Fd\bar{3}m$ (227) & 0.0 & camd-49983 & \ce{Mg3(BRh2)2} & $P2/m$ (10) & 0.0 & camd-110591 \\
\ce{Mg3(CoP)2} & $Cmcm$ (63) & 0.0 & camd-48431 & \ce{Mg3B3Pt4} & $I\bar{4}3d$ (220) & 0.0 & camd-1110 \\
\ce{MgB2Ir3} & $P6/mmm$ (191) & 0.0 & camd-10521 & \ce{MgB4Mo} & $P6/mmm$ (191) & 0.0 & camd-142130 \\
\ce{MgBPd3} & $I4/mmm$ (139) & 0.0 & camd-101635 & \ce{MgBPt3} & $P6/mmm$ (191) & 0.0 & camd-4775 \\
\ce{MgBRh} & $P6_{222}$ (180) & 0.0 & camd-101164 & \ce{MgFeP} & $P4/nmm$ (129) & 0.0 & camd-51271 \\
\ce{MgNbB4} & $P6/mmm$ (191) & 0.0 & camd-80730 & \ce{MgPIr} & $P2_{1}/c$ (14) & 0.0 & camd-70621 \\
\ce{MgPRh} & $Pnma$ (62) & 0.0 & camd-44069 & \ce{MgPdO3} & $R3c$ (161) & 0.0 & camd-53455 \\
\ce{MgTcO4} & $P2/c$ (13) & 0.0 & camd-49669 & \ce{MgTePd} & $F\bar{4}3m$ (216) & 0.0 & camd-71658 \\
\ce{Mn(AgI2)2} & $P\bar{4}2m$ (111) & 0.0 & camd-21673 & \ce{MnBr3} & $R\bar{3}$ (148) & 0.0 & camd-9335 \\
\ce{MnCl3} & $C2/c$ (15) & 0.0 & camd-34695 & \ce{MnCl4} & $P2/c$ (13) & 0.0 & camd-19543 \\
\ce{MnPIr} & $Pnma$ (62) & 0.0 & camd-56238 & \ce{MnPRu} & $Pnma$ (62) & 0.0 & camd-45867 \\
\ce{MoBr4} & $Cmc2_{1}$ (36) & 0.0 & camd-9366 & \ce{MoBr5} & $P\bar{1}$ (2) & 0.0 & camd-33146 \\
\ce{MoCl4} & $Cmc2_{1}$ (36) & 0.0 & camd-8968 & \ce{Na(CoP)2} & $I4/mmm$ (139) & 0.0 & camd-3436 \\
\ce{Na(Ni2P)2} & $P4_{2}/mnm$ (136) & 0.0 & camd-107223 & \ce{Na(PRh)2} & $I4/mmm$ (139) & 0.0 & camd-6739 \\
\ce{Na2AgP} & $Cmcm$ (63) & 0.0 & camd-71066 & \ce{Na2HfO3} & $Pmmn$ (59) & 0.0 & camd-48169 \\
\ce{Na2NiO2} & $Pbcn$ (60) & 0.0 & camd-38248 & \ce{Na2P2Ir} & $P4/mbm$ (127) & 0.0 & camd-45758 \\
\ce{Na2P2Rh} & $P4/mbm$ (127) & 0.0 & camd-51296 & \ce{Na2PdO2} & $Cmc2_{1}$ (36) & 0.0 & camd-152158 \\
\ce{Na3(NiP)4} & $I\bar{4}3m$ (217) & 0.0 & camd-45672 & \ce{Na3(PRh)4} & $I\bar{4}3m$ (217) & 0.0 & camd-25256 \\
\ce{Na3GaO3} & $P\bar{1}$ (2) & 0.0 & camd-95663 & \ce{Na3GaP2} & $Ibam$ (72) & 0.0 & camd-51244 \\
\ce{Na3TcO4} & $P\bar{4}3m$ (215) & 0.0 & camd-179579 & \ce{Na4RuO5} & $P\bar{1}$ (2) & 0.0 & camd-56731 \\
\ce{NaFeP} & $P4/nmm$ (129) & 0.0 & camd-34554 & \ce{NaNiP} & $P4/nmm$ (129) & 0.0 & camd-58372 \\
\ce{NaPIr} & $Cmce$ (64) & 0.0 & camd-60707 & \ce{NaRuO3} & $Pnna$ (52) & 0.0 & camd-141841 \\
\ce{NaTcO3} & $Pnma$ (62) & 0.0 & camd-148658 & \ce{NaTcO4} & $I4_{1}md$ (109) & 0.0 & camd-61018 \\
\ce{NaTi2O4} & $Cmcm$ (63) & 0.0 & camd-75221 & \ce{NaZnP} & $P6_{3}/mmc$ (194) & 0.0 & camd-9134 \\
\ce{NiCl3} & $R\bar{3}$ (148) & 0.0 & camd-8363 & \ce{NiP3Ru4} & $P6_{3}mc$ (186) & 0.0 & camd-79749 \\
\ce{NiPRu} & $P4/nmm$ (129) & 0.0 & camd-173835 & \ce{OsBr4} & $Pbam$ (55) & 0.0 & camd-14746 \\
\ce{OsBr5} & $P\bar{1}$ (2) & 0.0 & camd-14755 & \ce{OsCl5} & $P\bar{1}$ (2) & 0.0 & camd-4502 \\
\ce{OsCl6} & $C2/c$ (15) & 0.0 & camd-3713 & \ce{PdCl4} & $P2/c$ (13) & 0.0 & camd-35881 \\
\ce{PdF2} & $P3_{1}21$ (152) & 0.0 & camd-7057 & \ce{PtBr4} & $P2/c$ (13) & 0.0 & camd-7947 \\
\ce{PtCl4} & $P2/c$ (13) & 0.0 & camd-24455 & \ce{PtCl5} & $P\bar{1}$ (2) & 0.0 & camd-19949 \\
\ce{Rb(FeP)2} & $Immm$ (71) & 0.0 & camd-30406 & \ce{Rb2PdO2} & $Immm$ (71) & 0.0 & camd-65409 \\
\ce{Rb2TcO4} & $P\bar{3}m1$ (164) & 0.0 & camd-46470 & \ce{Rb3NbO4} & $I\bar{4}3m$ (217) & 0.0 & camd-75796 \\
\ce{Rb3PtO4} & $I4/mmm$ (139) & 0.0 & camd-161964 & \ce{Rb3RuO4} & $I\bar{4}2m$ (121) & 0.0 & camd-159987 \\
\ce{Rb3VO4} & $Pmn2_{1}$ (31) & 0.0 & camd-163751 & \ce{Rb4CrO4} & $I\bar{4}2m$ (121) & 0.0 & camd-64291 \\
\ce{RbAlO2} & $Pna2_{1}$ (33) & 0.0 & camd-84589 & \ce{RbInO2} & $Pna2_{1}$ (33) & 0.0 & camd-53601 \\
\ce{ReBr4} & $P2/c$ (13) & 0.0 & camd-1372 & \ce{ReBr5} & $P\bar{1}$ (2) & 0.0 & camd-1375 \\
\ce{ReBr6} & $C2/c$ (15) & 0.0 & camd-1381 & \ce{ReCl5} & $P\bar{1}$ (2) & 0.0 & camd-26019 \\
\ce{ReCl6} & $C2/c$ (15) & 0.0 & camd-21923 & \ce{RhI3} & $C2/c$ (15) & 0.0 & camd-14168 \\
\ce{RuCl5} & $P\bar{1}$ (2) & 0.0 & camd-14626 & \ce{RuF6} & $C2/c$ (15) & 0.0 & camd-31537 \\
\ce{RuI3} & $P\bar{3}1c$ (163) & 0.0 & camd-842 & \ce{SbBr3} & $R\bar{3}$ (148) & 0.0 & camd-8301 \\
\ce{SbF5} & $P\bar{1}$ (2) & 0.0 & camd-1420 & \ce{Sc(FeP)2} & $I4/mmm$ (139) & 0.0 & camd-34422 \\
\ce{Sc3Te2} & $Pnma$ (62) & 0.0 & camd-186132 & \ce{Sc5S6} & $C2/m$ (12) & 0.0 & camd-46308 \\
\ce{ScBr2} & $Fmm2$ (42) & 0.0 & camd-24287 & \ce{ScBr3} & $R\bar{3}$ (148) & 0.0 & camd-13707 \\
\ce{ScFeP} & $Pnma$ (62) & 0.0 & camd-22617 & \ce{ScI3} & $P\bar{3}1c$ (163) & 0.0 & camd-32084 \\
\ce{ScS} & $P6_{3}/mmc$ (194) & 0.0 & camd-64614 & \ce{ScSb} & $P6_{3}/mmc$ (194) & 0.0 & camd-84347 \\
\ce{ScSe} & $P6_{3}/mmc$ (194) & 0.0 & camd-42705 & \ce{SiI4} & $I\bar{4}2m$ (121) & 0.0 & camd-21450 \\
\ce{SnI2} & $P\bar{1}$ (2) & 0.0 & camd-37079 & \ce{Sr(GaP)2} & $P2/m$ (10) & 0.0 & camd-38528 \\
\ce{Sr(IrO3)2} & $Imma$ (74) & 0.0 & camd-35983 & \ce{Sr(MnP)2} & $I4/mmm$ (139) & 0.0 & camd-70694 \\
\ce{Sr(PRh)2} & $P3_{1}21$ (154) & 0.0 & camd-127317 & \ce{Sr(PdO2)3} & $Cmmm$ (65) & 0.0 & camd-32716 \\
\ce{Sr(PtO2)2} & $I4_{1}/a$ (88) & 0.0 & camd-17352 & \ce{Sr2ZnP2} & $Ibam$ (72) & 0.0 & camd-57586 \\
\ce{Sr3MoO6} & $P2_{1}/c$ (14) & 0.0 & camd-91729 & \ce{Sr3PRh2} & $R\bar{3}c$ (167) & 0.0 & camd-70878 \\
\ce{SrHfP2} & $R\bar{3}m$ (166) & 0.0 & camd-78431 & \ce{SrI2} & $Fd\bar{3}m$ (227) & 0.0 & camd-19307 \\
\ce{SrIrO3} & $Cmcm$ (63) & 0.0 & camd-25423 & \ce{SrNiP} & $P6_{3}/mmc$ (194) & 0.0 & camd-179612 \\
\ce{SrPRh} & $P2_{1}/c$ (14) & 0.0 & camd-56706 & \ce{SrPtO3} & $Pnma$ (62) & 0.0 & camd-12276 \\
\ce{SrV2O5} & $Pmmn$ (59) & 0.0 & camd-55388 & \ce{TiBr2} & $P\bar{6}m2$ (187) & 0.0 & camd-25287 \\
\ce{TiCl2} & $Fmm2$ (42) & 0.0 & camd-27891 & \ce{TiI4} & $P\bar{4}3m$ (215) & 0.0 & camd-2626 \\
\ce{TlF3} & $Im\bar{3}$ (204) & 0.0 & camd-17666 & \ce{VBr3} & $C2/c$ (15) & 0.0 & camd-33441 \\
\ce{VBr4} & $P\bar{4}3m$ (215) & 0.0 & camd-32457 & \ce{VCl3} & $P\bar{3}1m$ (162) & 0.0 & camd-23290 \\
\ce{VCl5} & $P6_{3}/mmc$ (194) & 0.0 & camd-36465 & \ce{VFe4} & $P4_{2}/m$ (84) & 0.0 & camd-9747 \\
\ce{VI3} & $P\bar{3}1c$ (163) & 0.0 & camd-18024 & \ce{VI4} & $P\bar{4}3m$ (215) & 0.0 & camd-8400 \\
\ce{Y(FeP)2} & $I4/mmm$ (139) & 0.0 & camd-26369 & \ce{Y2Zn2Te} & $P\bar{6}m2$ (187) & 0.0 & camd-80940 \\
\ce{Y3Te4} & $Pbam$ (55) & 0.0 & camd-104993 & \ce{Y4AlSe3} & $Pm\bar{3}m$ (221) & 0.0 & camd-58835 \\
\ce{Y4ZnSe3} & $Pm\bar{3}m$ (221) & 0.0 & camd-163232 & \ce{YBr3} & $P\bar{3}1m$ (162) & 0.0 & camd-14814 \\
\ce{YTe} & $P6_{3}/mmc$ (194) & 0.0 & camd-69171 & \ce{Zn2SbRh3} & $P\bar{3}m1$ (164) & 0.0 & camd-63378 \\
\ce{ZnBIr2} & $P6_{3}/mmc$ (194) & 0.0 & camd-81812 & \ce{ZrBr2} & $P\bar{6}m2$ (187) & 0.0 & camd-27793 \\
\ce{ZrCu2Se3} & $P2_{1}/m$ (11) & 0.0 & camd-172824 & \ce{Na2HfO3} & $Immm$ (71) & 0.0 & camd-40212 \\
\ce{TlI7} & $P\bar{1}$ (2) & 0.1 & camd-21398 & \ce{KNiP} & $P4/nmm$ (129) & 0.1 & camd-84441 \\
\ce{MnBr4} & $P2/c$ (13) & 0.2 & camd-28349 & \ce{RhF3} & $Fd\bar{3}m$ (227) & 0.3 & camd-24088 \\
\ce{CaCdO2} & $Pmmn$ (59) & 0.4 & camd-124516 & \ce{AlF3} & $Im\bar{3}$ (204) & 0.4 & camd-35349 \\
\ce{AuI} & $Cmce$ (64) & 0.4 & camd-1938 & \ce{CrF3} & $Fd\bar{3}m$ (227) & 0.4 & camd-708 \\
\ce{VCl4} & $P\bar{4}3m$ (215) & 0.5 & camd-3880 & \ce{SrTiO3} & $Cmcm$ (63) & 0.6 & camd-77955 \\
\ce{TlCl3} & $P\bar{3}1c$ (163) & 0.6 & camd-37503 & \ce{SrI2} & $P\bar{3}m1$ (164) & 0.6 & camd-4245 \\
\ce{BaIrO3} & $P6_{3}/mmc$ (194) & 0.7 & camd-19475 & \ce{Sr(NiP2)2} & $I4/mmm$ (139) & 0.7 & camd-171057 \\
\ce{Ca2AgP} & $P2_{1}/m$ (11) & 0.7 & camd-90894 & \ce{MgPIr} & $Pnma$ (62) & 0.8 & camd-39588 \\
\ce{KPIr} & $P2_{1}/c$ (14) & 0.8 & camd-68341 & \ce{TeF6} & $Im\bar{3}m$ (229) & 0.9 & camd-37559 \\
\ce{NbCl5} & $P\bar{1}$ (2) & 0.9 & camd-34513 & \ce{RhCl3} & $C2/c$ (15) & 0.9 & camd-8428 \\
\end{longtable}
\endgroup
\restoregeometry

\clearpage
\newgeometry{margin=1.8cm}            
\begingroup
\scriptsize
\setlength\tabcolsep{3pt}
\renewcommand{\arraystretch}{1.15}
\setlength{\LTcapwidth}{\textwidth}

\begin{longtable}{@{}llccl @{\hspace{1em}} llccl@{}}
\caption{The 166 dynamically stable materials identified by combining analysis of harmonic phonon spectra with finite-temperature molecular dynamics simulations. For each material, the table lists the chemical formula, space group, $E^{\mathrm{PBE}}_{\mathrm{hull}}$ and $E^{\mathrm{SCAN}}_{\mathrm{hull}}$ (meV/atom), and the CAMD database ID.}
\label{tab:dynamically_stable} \\
\toprule
& & \multicolumn{2}{c}{$E_{\mathrm{hull}}$ (meV/atom)} & &
& & \multicolumn{2}{c}{$E_{\mathrm{hull}}$ (meV/atom)} & \\
\cmidrule(lr){3-4} \cmidrule(lr){8-9}
Material & Space group & PBE & SCAN & Database ID &
Material & Space group & PBE & SCAN & Database ID \\
\midrule
\endfirsthead

\multicolumn{10}{c}{{\bfseries \tablename\ \thetable{} -- continued from previous page}} \\
\toprule
& & \multicolumn{2}{c}{$E_{\mathrm{hull}}$ (meV/atom)} & &
& & \multicolumn{2}{c}{$E_{\mathrm{hull}}$ (meV/atom)} & \\
\cmidrule(lr){3-4} \cmidrule(lr){8-9}
Material & Space group & PBE & SCAN & Database ID &
Material & Space group & PBE & SCAN & Database ID \\
\midrule
\endhead

\midrule \multicolumn{10}{r}{{\scriptsize Continued on next page}} \\
\endfoot

\bottomrule
\endlastfoot
\ce{Ca2Cd3O5} & $C2/m$ (12) & 0.0 & 0.0 & camd-66539 & \ce{SrPtO3} & $Pnma$ (62) & 0.0 & 0.0 & camd-12276 \\
\ce{MnPIr} & $Pnma$ (62) & 0.0 & 0.0 & camd-56238 & \ce{VBr3} & $C2/c$ (15) & 0.0 & 0.0 & camd-33441 \\
\ce{NaNiP} & $P4/nmm$ (129) & 0.0 & 0.0 & camd-58372 & \ce{VFe4} & $P4_{2}/m$ (84) & 0.0 & 0.0 & camd-9747 \\
\ce{Mg(TcB)2} & $Cmmm$ (65) & 0.0 & 0.0 & camd-37677 & \ce{VI3} & $P\bar{3}1c$ (163) & 0.0 & 0.0 & camd-18024 \\
\ce{CaRuO4} & $Pbcn$ (60) & 0.0 & 0.0 & camd-47329 & \ce{Y2Zn2Te} & $P\bar{6}m2$ (187) & 0.0 & 0.0 & camd-80940 \\
\ce{CaTcO4} & $P2/c$ (13) & 0.0 & 0.0 & camd-128896 & \ce{Y4AlSe3} & $Pm\bar{3}m$ (221) & 0.0 & 0.0 & camd-58835 \\
\ce{MgTcO4} & $P2/c$ (13) & 0.0 & 0.0 & camd-49669 & \ce{Zn2SbRh3} & $P\bar{3}m1$ (164) & 0.0 & 0.0 & camd-63378 \\
\ce{Na(CoP)2} & $I4/mmm$ (139) & 0.0 & 0.0 & camd-3436 & \ce{CaCdO2} & $Pmmn$ (59) & 0.4 & 0.0 & camd-124516 \\
\ce{AgCl2} & $P2_{1}/c$ (14) & 0.0 & 0.0 & camd-3286 & \ce{Sr(NiP2)2} & $I4/mmm$ (139) & 0.7 & 0.0 & camd-171057 \\
\ce{Al(TcB)2} & $Cmmm$ (65) & 0.0 & 0.0 & camd-43082 & \ce{MgPIr} & $Pnma$ (62) & 0.8 & 0.0 & camd-39588 \\
\ce{Ba(Cu2P)2} & $Cmmm$ (65) & 0.0 & 0.0 & camd-44902 & \ce{KPIr} & $P2_{1}/c$ (14) & 0.8 & 0.0 & camd-68341 \\
\ce{Ba(IrO3)2} & $P\bar{3}1m$ (162) & 0.0 & 0.0 & camd-15889 & \ce{Fe2NiP2Ru} & $Pmc2_{1}$ (26) & 0.0 & 0.0 & camd-89058 \\
\ce{Ba(ZnP)2} & $Pnma$ (62) & 0.0 & 0.0 & camd-141216 & \ce{Li(Ni2P)2} & $P4_{2}/mnm$ (136) & 0.0 & 0.0 & camd-267 \\
\ce{Ba2TcO4} & $I4/mmm$ (139) & 0.0 & 0.0 & camd-30164 & \ce{Mg(ZnP)2} & $P\bar{3}m1$ (164) & 0.0 & 0.0 & camd-56520 \\
\ce{BaIn2O4} & $P2_{1}/m$ (11) & 0.0 & 0.0 & camd-21023 & \ce{Ca(PdO2)3} & $Cmmm$ (65) & 0.0 & 0.0 & camd-65605 \\
\ce{BaPdO3} & $C2/m$ (12) & 0.0 & 0.0 & camd-21317 & \ce{Sr(PdO2)3} & $Cmmm$ (65) & 0.0 & 0.0 & camd-32716 \\
\ce{BaPtO3} & $C2/c$ (15) & 0.0 & 0.0 & camd-12734 & \ce{SrNiP} & $P6_{3}/mmc$ (194) & 0.0 & 0.0 & camd-179612 \\
\ce{BiCl3} & $P\bar{3}1c$ (163) & 0.0 & 0.0 & camd-29297 & \ce{BaNiP} & $Pnma$ (62) & 0.0 & 0.0 & camd-43939 \\
\ce{Ca(IrO3)2} & $P\bar{3}1m$ (162) & 0.0 & 0.0 & camd-79291 & \ce{Fe2CoP2Ru} & $Pmc2_{1}$ (26) & 0.0 & 0.2 & camd-111559 \\
\ce{Ca2Cu3P4} & $Ibam$ (72) & 0.0 & 0.0 & camd-170835 & \ce{ReCl6} & $C2/c$ (15) & 0.0 & 0.4 & camd-21923 \\
\ce{Ca3BiP} & $Pnma$ (62) & 0.0 & 0.0 & camd-12078 & \ce{CaHf2P3} & $P\bar{3}m1$ (164) & 0.0 & 0.5 & camd-68524 \\
\ce{Ca3Cd2O5} & $C2/m$ (12) & 0.0 & 0.0 & camd-76036 & \ce{NbCl5} & $P\bar{1}$ (2) & 0.9 & 0.7 & camd-34513 \\
\ce{Ca3RuO6} & $R\bar{3}$ (148) & 0.0 & 0.0 & camd-81624 & \ce{RhI3} & $C2/c$ (15) & 0.0 & 0.7 & camd-14168 \\
\ce{CaCd5O6} & $C2/m$ (12) & 0.0 & 0.0 & camd-74563 & \ce{ReCl5} & $P\bar{1}$ (2) & 0.0 & 0.7 & camd-26019 \\
\ce{CaNiP} & $P\bar{6}m2$ (187) & 0.0 & 0.0 & camd-40483 & \ce{RhCl3} & $C2/c$ (15) & 0.9 & 0.7 & camd-8428 \\
\ce{CaPIr} & $P2_{13}$ (198) & 0.0 & 0.0 & camd-56579 & \ce{BaIrO3} & $P6_{3}/mmc$ (194) & 0.7 & 0.8 & camd-19475 \\
\ce{CaPRh} & $P2_{1}/c$ (14) & 0.0 & 0.0 & camd-43504 & \ce{Na3(PRh)4} & $I\bar{4}3m$ (217) & 0.0 & 1.4 & camd-25256 \\
\ce{CoBr3} & $R\bar{3}$ (148) & 0.0 & 0.0 & camd-6836 & \ce{MgBPd3} & $I4/mmm$ (139) & 0.0 & 1.8 & camd-101635 \\
\ce{CoCl2} & $I\bar{4}3m$ (217) & 0.0 & 0.0 & camd-19238 & \ce{Mg3(BRh2)2} & $P2/m$ (10) & 0.0 & 2.0 & camd-110591 \\
\ce{CoCl3} & $P\bar{3}1m$ (162) & 0.0 & 0.0 & camd-21981 & \ce{SrI2} & $P\bar{3}m1$ (164) & 0.6 & 2.7 & camd-4245 \\
\ce{CoI2} & $P3_{1}21$ (154) & 0.0 & 0.0 & camd-1649 & \ce{AlF3} & $Im\bar{3}$ (204) & 0.4 & 2.9 & camd-35349 \\
\ce{CsPtO3} & $Cmcm$ (63) & 0.0 & 0.0 & camd-37880 & \ce{AgF3} & $Im\bar{3}$ (204) & 0.0 & 5.1 & camd-36738 \\
\ce{Cu3B3Pt4} & $I\bar{4}3d$ (220) & 0.0 & 0.0 & camd-64822 & \ce{ZrCu2Se3} & $P2_{1}/m$ (11) & 0.0 & 5.2 & camd-172824 \\
\ce{Cu3TcSe4} & $P\bar{4}3m$ (215) & 0.0 & 0.0 & camd-176274 & \ce{BaBr2} & $Fm\bar{3}m$ (225) & 0.0 & 5.2 & camd-14412 \\
\ce{CuBPt2} & $Cmcm$ (63) & 0.0 & 0.0 & camd-108676 & \ce{LiCoP} & $P4/nmm$ (129) & 0.0 & 6.3 & camd-3466 \\
\ce{CuF3} & $Im\bar{3}$ (204) & 0.0 & 0.0 & camd-9764 & \ce{RbInO2} & $Pna2_{1}$ (33) & 0.0 & 6.5 & camd-53601 \\
\ce{CuTcSe4} & $P2/m$ (10) & 0.0 & 0.0 & camd-89283 & \ce{KFeP} & $P4/nmm$ (129) & 0.0 & 7.0 & camd-12435 \\
\ce{InCl3} & $C2/c$ (15) & 0.0 & 0.0 & camd-22548 & \ce{Cu2SbRh3} & $P\bar{3}m1$ (164) & 0.0 & 7.9 & camd-85589 \\
\ce{K(CuP2)2} & $I4/mmm$ (139) & 0.0 & 0.0 & camd-152270 & \ce{Na3GaO3} & $P\bar{1}$ (2) & 0.0 & 8.0 & camd-95663 \\
\ce{K2HfS3} & $C2/m$ (12) & 0.0 & 0.0 & camd-34482 & \ce{SrI2} & $Fd\bar{3}m$ (227) & 0.0 & 8.0 & camd-19307 \\
\ce{K3Al2P3} & $P2_{1}/m$ (11) & 0.0 & 0.0 & camd-49812 & \ce{MgPRh} & $Pnma$ (62) & 0.0 & 9.2 & camd-44069 \\
\ce{Li(CoP)2} & $I4/mmm$ (139) & 0.0 & 0.0 & camd-35644 & \ce{Ca3Ag3P} & $Cmcm$ (63) & 0.0 & 9.2 & camd-67218 \\
\ce{Li3AgO3} & $P4_{2}/mnm$ (136) & 0.0 & 0.0 & camd-115567 & \ce{K3InO3} & $P\bar{1}$ (2) & 0.0 & 10.2 & camd-41331 \\
\ce{Li3RuO4} & $C2/c$ (15) & 0.0 & 0.0 & camd-75839 & \ce{SrV2O5} & $Pmmn$ (59) & 0.0 & 12.9 & camd-55388 \\
\ce{LiAgO2} & $P4_{2}/mmc$ (131) & 0.0 & 0.0 & camd-159085 & \ce{BaI2} & $Fm\bar{3}m$ (225) & 0.0 & 14.1 & camd-20756 \\
\ce{LiCuO2} & $P2_{1}/c$ (14) & 0.0 & 0.0 & camd-66501 & \ce{NaFeP} & $P4/nmm$ (129) & 0.0 & 16.4 & camd-34554 \\
\ce{Mg(RuO3)2} & $P4_{2}/mnm$ (136) & 0.0 & 0.0 & camd-53651 & \ce{Ca(MnP)2} & $I4/mmm$ (139) & 0.0 & 16.5 & camd-94778 \\
\ce{Mg2BIr2} & $Pbam$ (55) & 0.0 & 0.0 & camd-22036 & \ce{CdI2} & $I\bar{4}3m$ (217) & 0.0 & 19.5 & camd-18153 \\
\ce{Mg2Ni3P} & $P6_{3}/mmc$ (194) & 0.0 & 0.0 & camd-134578 & \ce{MgPIr} & $P2_{1}/c$ (14) & 0.0 & 20.0 & camd-70621 \\
\ce{Mg2SbRh3} & $P\bar{3}m1$ (164) & 0.0 & 0.0 & camd-185530 & \ce{GaF3} & $Fd\bar{3}m$ (227) & 0.0 & 21.1 & camd-19356 \\
\ce{Mg3B3Pt4} & $I\bar{4}3d$ (220) & 0.0 & 0.0 & camd-1110 & \ce{ZnBIr2} & $P6_{3}/mmc$ (194) & 0.0 & 21.2 & camd-81812 \\
\ce{MgBRh} & $P6_{222}$ (180) & 0.0 & 0.0 & camd-101164 & \ce{CrF3} & $Fd\bar{3}m$ (227) & 0.4 & 24.1 & camd-708 \\
\ce{MgNbB4} & $P6/mmm$ (191) & 0.0 & 0.0 & camd-80730 & \ce{IrCl5} & $P\bar{1}$ (2) & 0.0 & 26.0 & camd-33422 \\
\ce{MgPdO3} & $R3c$ (161) & 0.0 & 0.0 & camd-53455 & \ce{VCl3} & $P\bar{3}1m$ (162) & 0.0 & 27.0 & camd-23290 \\
\ce{MgTePd} & $F\bar{4}3m$ (216) & 0.0 & 0.0 & camd-71658 & \ce{Ca2IrO4} & $Cmmm$ (65) & 0.0 & 29.6 & camd-170316 \\
\ce{MnBr3} & $R\bar{3}$ (148) & 0.0 & 0.0 & camd-9335 & \ce{ScFeP} & $Pnma$ (62) & 0.0 & 29.6 & camd-22617 \\
\ce{MnCl3} & $C2/c$ (15) & 0.0 & 0.0 & camd-34695 & \ce{OsCl6} & $C2/c$ (15) & 0.0 & 30.4 & camd-3713 \\
\ce{Na(Ni2P)2} & $P4_{2}/mnm$ (136) & 0.0 & 0.0 & camd-107223 & \ce{Ca2RhO4} & $Cmmm$ (65) & 0.0 & 30.5 & camd-46457 \\
\ce{Na(PRh)2} & $I4/mmm$ (139) & 0.0 & 0.0 & camd-6739 & \ce{IrF3} & $Fd\bar{3}m$ (227) & 0.0 & 32.2 & camd-17173 \\
\ce{Na2AgP} & $Cmcm$ (63) & 0.0 & 0.0 & camd-71066 & \ce{RhF3} & $Fd\bar{3}m$ (227) & 0.3 & 35.3 & camd-24088 \\
\ce{Na3(NiP)4} & $I\bar{4}3m$ (217) & 0.0 & 0.0 & camd-45672 & \ce{Ba2RuO5} & $P4/mmm$ (123) & 0.0 & 35.8 & camd-4628 \\
\ce{Na3TcO4} & $P\bar{4}3m$ (215) & 0.0 & 0.0 & camd-179579 & \ce{SbBr3} & $R\bar{3}$ (148) & 0.0 & 39.7 & camd-8301 \\
\ce{NaTcO3} & $Pnma$ (62) & 0.0 & 0.0 & camd-148658 & \ce{NaPIr} & $Cmce$ (64) & 0.0 & 40.0 & camd-60707 \\
\ce{NaZnP} & $P6_{3}/mmc$ (194) & 0.0 & 0.0 & camd-9134 & \ce{Ca3AlP3} & $P\bar{1}$ (2) & 0.0 & 42.5 & camd-43464 \\
\ce{OsCl5} & $P\bar{1}$ (2) & 0.0 & 0.0 & camd-4502 & \ce{CaBr2} & $P\bar{1}$ (2) & 0.0 & 42.8 & camd-20618 \\
\ce{PtBr4} & $P2/c$ (13) & 0.0 & 0.0 & camd-7947 & \ce{MnPRu} & $Pnma$ (62) & 0.0 & 43.4 & camd-45867 \\
\ce{Rb(FeP)2} & $Immm$ (71) & 0.0 & 0.0 & camd-30406 & \ce{CrSbS4} & $C2/m$ (12) & 0.0 & 45.9 & camd-162358 \\
\ce{RbAlO2} & $Pna2_{1}$ (33) & 0.0 & 0.0 & camd-84589 & \ce{Ca4AlP} & $P2/c$ (13) & 0.0 & 48.5 & camd-50716 \\
\ce{RuCl5} & $P\bar{1}$ (2) & 0.0 & 0.0 & camd-14626 & \ce{Ca3Al3P} & $P2_{1}/m$ (11) & 0.0 & 48.9 & camd-80123 \\
\ce{RuI3} & $P\bar{3}1c$ (163) & 0.0 & 0.0 & camd-842 & \ce{YBr3} & $P\bar{3}1m$ (162) & 0.0 & 53.5 & camd-14814 \\
\ce{Sc5S6} & $C2/m$ (12) & 0.0 & 0.0 & camd-46308 & \ce{Ca2AgP} & $P2_{1}/m$ (11) & 0.7 & 53.8 & camd-90894 \\
\ce{ScBr3} & $R\bar{3}$ (148) & 0.0 & 0.0 & camd-13707 & \ce{MgFeP} & $P4/nmm$ (129) & 0.0 & 55.8 & camd-51271 \\
\ce{ScI3} & $P\bar{3}1c$ (163) & 0.0 & 0.0 & camd-32084 & \ce{CaCoP} & $P4/nmm$ (129) & 0.0 & 60.4 & camd-53623 \\
\ce{ScS} & $P6_{3}/mmc$ (194) & 0.0 & 0.0 & camd-64614 & \ce{PdF2} & $P3_{1}21$ (152) & 0.0 & 68.8 & camd-7057 \\
\ce{ScSb} & $P6_{3}/mmc$ (194) & 0.0 & 0.0 & camd-84347 & \ce{Mg3(CoP)2} & $Cmcm$ (63) & 0.0 & 75.2 & camd-48431 \\
\ce{ScSe} & $P6_{3}/mmc$ (194) & 0.0 & 0.0 & camd-42705 & \ce{Sc(FeP)2} & $I4/mmm$ (139) & 0.0 & 78.2 & camd-34422 \\
\ce{SnI2} & $P\bar{1}$ (2) & 0.0 & 0.0 & camd-37079 & \ce{Ca2ZrP2} & $P\bar{3}m1$ (164) & 0.0 & 81.7 & camd-43507 \\
\ce{Sr(GaP)2} & $P2/m$ (10) & 0.0 & 0.0 & camd-38528 & \ce{Ca2ZnP} & $Amm2$ (38) & 0.0 & 87.0 & camd-92647 \\
\ce{Sr(MnP)2} & $I4/mmm$ (139) & 0.0 & 0.0 & camd-70694 & \ce{Y(FeP)2} & $I4/mmm$ (139) & 0.0 & 100.8 & camd-26369 \\
\ce{Sr2ZnP2} & $Ibam$ (72) & 0.0 & 0.0 & camd-57586 & \ce{NiCl3} & $R\bar{3}$ (148) & 0.0 & 108.9 & camd-8363 \\
\ce{Sr3PRh2} & $R\bar{3}c$ (167) & 0.0 & 0.0 & camd-70878 & \ce{CaFeP} & $P4/nmm$ (129) & 0.0 & 112.1 & camd-29355 \\
\ce{SrIrO3} & $Cmcm$ (63) & 0.0 & 0.0 & camd-25423 & \ce{BaMnO2} & $Pmmn$ (59) & 0.0 & 210.2 & camd-26647 \\
\ce{SrPRh} & $P2_{1}/c$ (14) & 0.0 & 0.0 & camd-56706 & \ce{BaMn2O4} & $P2_{1}/m$ (11) & 0.0 & 258.3 & camd-4135 \\
\end{longtable}
\endgroup
\restoregeometry                      

\clearpage
\newgeometry{margin=1.8cm}
\begingroup
\footnotesize
\setlength\tabcolsep{3pt}
\renewcommand{\arraystretch}{1.05}
\setlength{\LTcapwidth}{\textwidth}

\begin{longtable}{@{}llccl @{\hspace{1em}} llccl@{}}
\caption{Synthesizability analysis of the 109 stable materials. For each material, the table lists the chemical formula, space group, decomposition enthalpy $\Delta H_{\mathrm{d}}$ (meV/atom), composite synthesizability score ($S$), and CAMD database ID. Materials are sorted in descending order of $S$; the 25 highest‑scoring $S$ values are shown in bold.}
\label{tab:S_scores} \\
\toprule
Material & Space group & $\Delta H_{\mathrm{d}}$ & $S$ score & Database ID &
Material & Space group & $\Delta H_{\mathrm{d}}$ & $S$ score & Database ID \\
\midrule
\endfirsthead

\multicolumn{10}{c}{{\bfseries \tablename\ \thetable{} -- continued from previous page}} \\
\toprule
Material & Space group & $\Delta H_{\mathrm{d}}$ & $S$ score & Database ID &
Material & Space group & $\Delta H_{\mathrm{d}}$ & $S$ score & Database ID \\
\midrule
\endhead

\midrule \multicolumn{10}{r}{{\footnotesize Continued on next page}} \\
\endfoot

\bottomrule
\endlastfoot
\ce{LiCuO2} & $P2_{1}/c$ (14) & -20.3 & \textbf{1.000} & camd-66501 & \ce{CaTcO4} & $P2/c$ (13) & -127.1 & \textbf{0.826} & camd-128896 \\
\ce{InCl3} & $C2/c$ (15) & -118.7 & \textbf{0.992} & camd-22548 & \ce{CaRuO4} & $Pbcn$ (60) & -29.8 & \textbf{0.812} & camd-47329 \\
\ce{CuF3} & $Im\bar{3}$ (204) & -311.0 & \textbf{0.971} & camd-9764 & \ce{Sc5S6} & $C2/m$ (12) & -53.6 & \textbf{0.808} & camd-46308 \\
\ce{CoCl3} & $P\bar{3}1m$ (162) & -310.8 & \textbf{0.971} & camd-21981 & \ce{SnI2} & $P\bar{1}$ (2) & -32.6 & \textbf{0.807} & camd-37079 \\
\ce{BaPdO3} & $C2/m$ (12) & -137.4 & \textbf{0.957} & camd-21317 & \ce{SrPRh} & $P2_{1}/c$ (14) & -105.1 & \textbf{0.807} & camd-56706 \\
\ce{Ca(IrO3)2} & $P\bar{3}1m$ (162) & -114.1 & \textbf{0.951} & camd-79291 & \ce{BaNiP} & $Pnma$ (62) & -35.9 & \textbf{0.802} & camd-43939 \\
\ce{BaPtO3} & $C2/c$ (15) & -106.0 & \textbf{0.930} & camd-12734 & \ce{Ca(PdO2)3} & $Cmmm$ (65) & -116.7 & 0.785 & camd-65605 \\
\ce{AgCl2} & $P2_{1}/c$ (14) & -111.8 & \textbf{0.903} & camd-3286 & \ce{CoBr3} & $R\bar{3}$ (148) & -116.4 & 0.785 & camd-6836 \\
\ce{Ba(IrO3)2} & $P\bar{3}1m$ (162) & -35.8 & \textbf{0.898} & camd-15889 & \ce{ScSe} & $P6_{3}/mmc$ (194) & -37.7 & 0.772 & camd-42705 \\
\ce{CaPIr} & $P2_{1}3$ (198) & -127.3 & \textbf{0.877} & camd-56579 & \ce{LiAgO2} & $P4_{2}/mmc$ (131) & -30.9 & 0.770 & camd-159085 \\
\ce{BaIn2O4} & $P2_{1}/m$ (11) & -19.4 & \textbf{0.877} & camd-21023 & \ce{CaPRh} & $P2_{1}/c$ (14) & -53.0 & 0.735 & camd-43504 \\
\ce{Sr(PdO2)3} & $Cmmm$ (65) & -80.1 & \textbf{0.868} & camd-32716 & \ce{VBr3} & $C2/c$ (15) & -70.0 & 0.732 & camd-33441 \\
\ce{RuCl5} & $P\bar{1}$ (2) & -115.1 & \textbf{0.867} & camd-14626 & \ce{MnBr3} & $R\bar{3}$ (148) & -28.2 & 0.721 & camd-9335 \\
\ce{SrNiP} & $P6_{3}/mmc$ (194) & -50.1 & \textbf{0.866} & camd-179612 & \ce{Mg2BIr2} & $Pbam$ (55) & -55.5 & 0.708 & camd-22036 \\
\ce{Li3AgO3} & $P4_{2}/mnm$ (136) & -72.5 & \textbf{0.858} & camd-115567 & \ce{BiCl3} & $P\bar{3}1c$ (163) & -54.7 & 0.706 & camd-29297 \\
\ce{Ca3RuO6} & $R\bar{3}$ (148) & -45.3 & \textbf{0.855} & camd-81624 & \ce{CoCl2} & $I\bar{4}3m$ (217) & -23.5 & 0.703 & camd-19238 \\
\ce{SrIrO3} & $Cmcm$ (63) & -17.2 & \textbf{0.841} & camd-25423 & \ce{ScI3} & $P\bar{3}1c$ (163) & -73.8 & 0.698 & camd-32084 \\
\ce{Ba(Cu2P)2} & $Cmmm$ (65) & -26.1 & \textbf{0.835} & camd-44902 & \ce{ScS} & $P6_{3}/mmc$ (194) & -16.8 & 0.690 & camd-64614 \\
\ce{CaNiP} & $P\bar{6}m2$ (187) & -137.9 & \textbf{0.835} & camd-40483 & \ce{Mg2SbRh3} & $P\bar{3}m1$ (164) & -109.7 & 0.689 & camd-185530 \\
\newpage
\ce{MgNbB4} & $P6/mmm$ (191) & -40.8 & 0.676 & camd-80730 & \ce{ReCl5} & $P\bar{1}$ (2) & -2.1 & 0.504 & camd-26019 \\
\ce{Sr2ZnP2} & $Ibam$ (72) & -58.5 & 0.674 & camd-57586 & \ce{Li3RuO4} & $C2/c$ (15) & -0.4 & 0.498 & camd-75839 \\
\ce{ScSb} & $P6_{3}/mmc$ (194) & -7.1 & 0.672 & camd-84347 & \ce{PtBr4} & $P2/c$ (13) & -15.4 & 0.488 & camd-7947 \\
\ce{Li(Ni2P)2} & $P4_{2}/mnm$ (136) & -16.8 & 0.669 & camd-267 & \ce{VI3} & $P\bar{3}1c$ (163) & -4.1 & 0.487 & camd-18024 \\
\ce{ScBr3} & $R\bar{3}$ (148) & -77.0 & 0.652 & camd-13707 & \ce{Sr(GaP)2} & $P2/m$ (10) & -15.0 & 0.485 & camd-38528 \\
\ce{K3Al2P3} & $P2_{1}/m$ (11) & -65.2 & 0.635 & camd-49812 & \ce{Ba2TcO4} & $I4/mmm$ (139) & -8.8 & 0.484 & camd-30164 \\
\ce{Ca2Cu3P4} & $Ibam$ (72) & -9.5 & 0.634 & camd-170835 & \ce{CsPtO3} & $Cmcm$ (63) & -83.4 & 0.466 & camd-37880 \\
\ce{K(CuP2)2} & $I4/mmm$ (139) & -26.5 & 0.632 & camd-152270 & \ce{CaCd5O6} & $C2/m$ (12) & -3.2 & 0.465 & camd-74563 \\
\ce{Mg(RuO3)2} & $P4_{2}/mnm$ (136) & -35.6 & 0.623 & camd-53651 & \ce{CuBPt2} & $Cmcm$ (63) & -11.9 & 0.462 & camd-108676 \\
\ce{CoI2} & $P3_{2}21$ (154) & -23.9 & 0.621 & camd-1649 & \ce{BaIrO3} & $P6_{3}/mmc$ (194) & 0.7 & 0.462 & camd-19475 \\
\ce{MnCl3} & $C2/c$ (15) & -52.1 & 0.611 & camd-34695 & \ce{Na2AgP} & $Cmcm$ (63) & -77.9 & 0.459 & camd-71066 \\
\ce{Li(CoP)2} & $I4/mmm$ (139) & -19.8 & 0.602 & camd-35644 & \ce{Ca2Cd3O5} & $C2/m$ (12) & -2.6 & 0.451 & camd-66539 \\
\ce{Na3TcO4} & $P\bar{4}3m$ (215) & -88.0 & 0.594 & camd-179579 & \ce{Ca3Cd2O5} & $C2/m$ (12) & -2.4 & 0.444 & camd-76036 \\
\ce{Sr(MnP)2} & $I4/mmm$ (139) & -43.2 & 0.592 & camd-70694 & \ce{ReCl6} & $C2/c$ (15) & -0.5 & 0.429 & camd-21923 \\
\ce{Mg2Ni3P} & $P6_{3}/mmc$ (194) & -5.2 & 0.558 & camd-134578 & \ce{VFe4} & $P4_{2}/m$ (84) & -0.1 & 0.413 & camd-9747 \\
\ce{Sr3PRh2} & $R\bar{3}c$ (167) & -8.9 & 0.557 & camd-70878 & \ce{CuTcSe4} & $P2/m$ (10) & -47.7 & 0.408 & camd-89283 \\
\ce{Ba(ZnP)2} & $Pnma$ (62) & -29.0 & 0.551 & camd-141216 & \ce{RbAlO2} & $Pna2_{1}$ (33) & -0.7 & 0.396 & camd-84589 \\
\ce{SrPtO3} & $Pnma$ (62) & -11.1 & 0.545 & camd-12276 & \ce{Na(CoP)2} & $I4/mmm$ (139) & -41.8 & 0.395 & camd-3436 \\
\ce{MgBRh} & $P6_{2}22$ (180) & -15.1 & 0.536 & camd-101164 & \ce{Fe2NiP2Ru} & $Pmc2_{1}$ (26) & -41.7 & 0.395 & camd-89058 \\
\newpage
\ce{Sr(NiP2)2} & $I4/mmm$ (139) & 0.7 & 0.394 & camd-171057 & \ce{MgTePd} & $F\bar{4}3m$ (216) & -12.0 & 0.269 & camd-71658 \\
\ce{Mg3B3Pt4} & $I\bar{4}3d$ (220) & -5.4 & 0.389 & camd-1110 & \ce{Na3(NiP)4} & $I\bar{4}3m$ (217) & -9.2 & 0.244 & camd-45672 \\
\ce{RuI3} & $P\bar{3}1c$ (163) & -5.4 & 0.389 & camd-842 & \ce{OsCl5} & $P\bar{1}$ (2) & 0.0 & 0.243 & camd-4502 \\
\ce{Na(PRh)2} & $I4/mmm$ (139) & -30.7 & 0.363 & camd-6739 & \ce{Y4AlSe3} & $Pm\bar{3}m$ (221) & -8.5 & 0.237 & camd-58835 \\
\ce{MgPdO3} & $R3c$ (161) & -28.9 & 0.357 & camd-53455 & \ce{Cu3TcSe4} & $P\bar{4}3m$ (215) & -7.7 & 0.227 & camd-176274 \\
\ce{Fe2CoP2Ru} & $Pmc2_{1}$ (26) & -28.0 & 0.354 & camd-111559 & \ce{RhCl3} & $C2/c$ (15) & 1.0 & 0.209 & camd-8428 \\
\ce{Mg(TcB)2} & $Cmmm$ (65) & -27.6 & 0.352 & camd-37677 & \ce{Cu3B3Pt4} & $I\bar{4}3d$ (220) & -0.1 & 0.199 & camd-64822 \\
\ce{Y2Zn2Te} & $P\bar{6}m2$ (187) & -27.5 & 0.352 & camd-80940 & \ce{Ca3BiP} & $Pnma$ (62) & 0.0 & 0.192 & camd-12078 \\
\ce{NbCl5} & $P\bar{1}$ (2) & 1.0 & 0.349 & camd-34513 & \ce{NaNiP} & $P4/nmm$ (129) & -5.2 & 0.191 & camd-58372 \\
\ce{Rb(FeP)2} & $Immm$ (71) & -25.0 & 0.342 & camd-30406 & \ce{CaHf2P3} & $P\bar{3}m1$ (164) & -3.0 & 0.145 & camd-68524 \\
\ce{NaTcO3} & $Pnma$ (62) & -6.7 & 0.335 & camd-148658 & \ce{MgTcO4} & $P2/c$ (13) & -0.2 & 0.141 & camd-49669 \\
\ce{Mg(ZnP)2} & $P\bar{3}m1$ (164) & -6.1 & 0.327 & camd-56520 & \ce{MnPIr} & $Pnma$ (62) & -2.7 & 0.137 & camd-56238 \\
\ce{Zn2SbRh3} & $P\bar{3}m1$ (164) & -20.8 & 0.324 & camd-63378 & \ce{MgPIr} & $Pnma$ (62) & 0.8 & 0.058 & camd-39588 \\
\ce{Al(TcB)2} & $Cmmm$ (65) & -15.7 & 0.296 & camd-43082 & \ce{KPIr} & $P2_{1}/c$ (14) & 0.8 & 0.056 & camd-68341 \\
\ce{RhI3} & $C2/c$ (15) & -0.6 & 0.295 & camd-14168 & \ce{Na(Ni2P)2} & $P4_{2}/mnm$ (136) & -0.5 & 0.041 & camd-107223 \\
\ce{NaZnP} & $P6_{3}/mmc$ (194) & -0.4 & 0.280 & camd-9134 & \ce{K2HfS3} & $C2/m$ (12) & 0.0 & 0.000 & camd-34482 \\
\ce{CaCdO2} & $Pmmn$ (59) & 0.4 & 0.278 & camd-124516 &  &  &  &  &  \\
\end{longtable}
\endgroup
\restoregeometry

\begin{figure}
    \centering
    \includegraphics[width=1\linewidth]{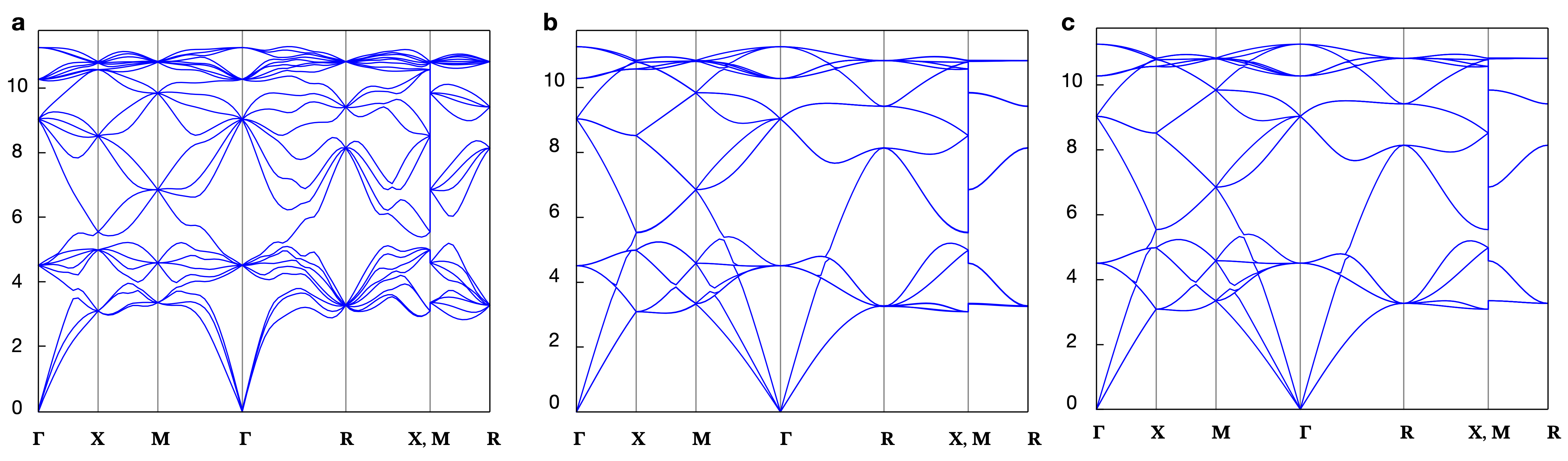}
    \caption{Phonon spectra of Si (mp-149) calculated using MACE for supercell sizes of (a) $2 \times 2 \times 2$, (b) $3 \times 3 \times 3$, and (c) $5 \times 5 \times 5$.}
    \label{fig:bandstructure-benchmark}
\end{figure}